\documentclass[sn-mathphys]{sn-jnl}% Math and Physical Sciences Reference Style
\jyear{2021}%

\theoremstyle{thmstyleone}%
\theoremstyle{thmstyletwo}%

\theoremstyle{thmstylethree}%

\usepackage{color}

\begin{document}

\title[Random walks in correlated diffusivity landscapes]{Random walks in correlated diffusivity landscapes}

%%=============================================================%%
%% Prefix	-> \pfx{Dr}
%% GivenName	-> \fnm{Joergen W.}
%% Particle	-> \spfx{van der} -> surname prefix
%% FamilyName	-> \sur{Ploeg}
%% Suffix	-> \sfx{IV}
%% NatureName	-> \tanm{Poet Laureate} -> Title after name
%% Degrees	-> \dgr{MSc, PhD}
%% \author*[1,2]{\pfx{Dr} \fnm{Joergen W.} \spfx{van der} \sur{Ploeg} \sfx{IV} \tanm{Poet Laureate} 
%%                 \dgr{MSc, PhD}}\email{iauthor@gmail.com}
%%=============================================================%%

\author*[1]{\fnm{Adrian} \sur{Pacheco-Pozo}}\email{adrian.pacheco@physik.hu-berlin.de}

\author[1,2]{\fnm{Igor M.} \sur{Sokolov}}\email{igor.sokolov@physik.hu-berlin.de}

\affil[1]{\orgdiv{Institut f\"ur Physik}, \orgname{Humboldt-Universit\"at zu Berlin}, \orgaddress{\street{Newtonstra{\ss}e 15}, \postcode{D-12489}, \city{Berlin},  \country{Germany}}}

\affil[2]{ \orgname{IRIS Adlershof}, \orgaddress{\street{Zum Gro{\ss}en Windkanal 2}, \postcode{D-12489}, \city{Berlin},  \country{Germany}}}

%%==================================%%
%% sample for unstructured abstract %%
%%==================================%%

\abstract{In recent years, several experiments highlighted a new type of diffusion anomaly, which was called Brownian yet non-Gaussian diffusion. In systems displaying this behavior, the mean squared displacement of the diffusing particles grows linearly in time, like in a normal diffusion, but the distribution of displacements is non-Gaussian. In situations when the convergence to Gaussian still takes place at longer times, the probability density of the displacements may show a persisting peak around the distribution's mode, and the pathway of convergence to the Gaussian is unusual. One of the theoretical models showing such a behavior corresponds to a disordered system with local diffusion coefficients slowly varying in space. While the standard pathway to Gaussian, as proposed by the Central Limit Theorem, would assume that the peak, under the corresponding rescaling, smoothens and lowers in course of the time; in the model discussed, the peak, under rescaling, narrows and stays sharp. In the present work, we discuss the nature of this peak. On a coarse-grained level, the motion of the particles in the diffusivity landscape is described by continuous time random walks with correlations between waiting times and positions. The peak is due to strong spatiotemporal correlations along the trajectories of diffusing particles. Destroying these correlations while keeping the temporal structure of the process intact leads to the decay of the peak. We also note that the correlated CTRW model reproducing serial correlations between the waiting times along the trajectory fails to quantitatively reproduce the shape of the peak even for the decorrelated motion, while being quite accurate in the wings of the PDF. This shows the importance of high-order temporal correlations for the peak's formation.}


%%\pacs[JEL Classification]{D8, H51}

%%\pacs[MSC Classification]{35A01, 65L10, 65L12, 65L20, 65L70}

\maketitle

%---------------------------------------------------------------
%Section: Introduction
%---------------------------------------------------------------

\section{Introduction}

The erratic motion of particles diffusing in a fluid medium (Brownian motion) has drawn considerable attention of scientists since Robert Brown first systematically investigated it \cite{Brown1828}. A. Einstein \cite{Einstein1905} was the first to propose a mathematical description of this type of motion (see \cite{Maiocchi1990} for a detailed historical account). Einstein, who essentially did not know about Brownian motion, found out, that such a phenomenon is an inavoidable consequence of the kinetic theory of heat, and closely connected it to diffusion. In this picture of what we now call \textit{normal diffusion}, the particles' motion possesses two important properties~\cite{Balakrishnan2019}: (i) The mean square displacement (MSD) of the 
particles from their initial position grows linearly in time,
\begin{equation}
\langle \mathbf{r}(t)^2 \rangle = 2 d D t
\label{eq:prop_1}
\end{equation}
(with $d$ being the dimension of space, and $D$ being the diffusion coefficient), and (ii) The probability density function (PDF) of the particles' displacements at a given time follows a Gaussian distribution 
\begin{equation}
p(\mathbf{r},t) = \frac{1}{(4 \pi D t)^{d/2}} \exp \left( -\frac{\mathbf{r}^2}{4 D t} \right).
\label{eq:Gauss}
\end{equation}
The properties (\ref{eq:prop_1}) and (\ref{eq:Gauss}) were tested in many experiments, and their confirmation laid a solid foundation to our understanding of the atomistic structure of matter \cite{Perrin1916}.  
The random walk approach used by Einstein assumed that one can approximate the particle's motion by a sequence of independent steps in random directions under the condition that the times
necessary to make a step are the same, and the displacement in a single step has a finite second moment. This approach was closely mirrored in many early experiments using stroboscopic measurements. Independently of Einstein, Smoluchowski \cite{Smoluchowski1906} presented a more formal mathematical description of the Brownian motion which lead to the same results as Einstein's, and set the ground to a new branch of probability theory concerning the diffusion processes \cite{Smoluchowski1916}. After Einstein and Smoluchowski, Langevin \cite{Langevin1908} proposed a new mathematical tool for the description of the particle's motion, the stochastic differential equation. 

The standard picture corresponds to the tracer's motion in a homogeneous, quiescent fluid. In the course of time, many deviations from this kind of behavior were found for other media. 
Numerous experiments on transport in complex media (disordered solids, rocks, biological media, etc.) showed that, instead of a linear time dependence as given by Eq.~(\ref{eq:prop_1}), the MSD often follows a power-law time-dependence $\langle \mathbf{r}(t)^2 \rangle \propto t^{\gamma}$, with $0 < \gamma < 1$ (subdiffusion) or $1 < \gamma < 2$ (superdiffusion). A system whose MSD shows such a time dependence is said to exhibit \textit{anomalous diffusion}. 
Depending on the specific case, different mathematical models have been proposed to describe this anomalous behavior by focusing on different aspects of the motion \cite{Metzler2000,Sokolov2012,Hofling2013,Krapf2015,Oliveira2019,Wang2022}. Some classical models are: the uncorrelated continuous time random walks (CTRW) with power-law waiting time distributions, the fractional Brownian motion, and L\'evy walks and L\'evy flights. The PDF in these models may or may not be Gaussian.

Several recent experiments \cite{Wang2009,Leptos2009,Kurtuldu2011,Wang2012,Skaug2013,Yu2013,He2013,Guan2014,Thorneywork2016,He2016,Acharya2017,Wagner2017,Chakraborty2019,Kwon2019,Chakraborty2020,Pastore2021,Pastore2022} reported a new type of diffusion in which the MSD grows linearly in time, like in the normal diffusion, yet the PDF of displacements shows considerable deviations from the Gaussian shape. 
Usually, the PDF of displacements is well-described by a Laplace (two-sided exponential) distribution. This behavior was called Brownian yet non-Gaussian (BnG) diffusion \cite{Wang2012}. Some of the corresponding systems show a crossover from the non-Gaussian distribution to a Gaussian one at long times \cite{Wang2009,Pastore2021}. In several cases \cite{Wang2009,Wang2012,Skaug2013,He2016,Wagner2017,Chakraborty2019,Kwon2019,Chakraborty2020,Pastore2021,Pastore2022}, for times at which the crossover takes place, the PDF presents a peak close to its mode. This peak resembles a part of the initial Laplace distribution, while the parts of the distribution further from its mode have already a more or less Gaussian shape.

Many of the systems in which the BnG diffusion is observed are pertinent to soft matter, and almost all of the experimental systems with BnG diffusion may show a great deal of spatial and temporal inhomogeneity, or disorder. Thus, the medium in which the particle moves may be spatially heterogeneous, 
or change in time. The properties of the diffusing tracer may change in time as well. 

Different assumptions about the heterogeneity involved lead to different classes of models which were proposed for the description of BnG diffusion. 
The most popular class corresponds to the diffusing diffusivity (DD) models, see e.g. \cite{Chubynsky2014,Jain2016,Chechkin2017,Lanoiselee2018_b,Slezak2018}. 
They assume slow random changes of the diffusion coefficient in time. The particular variant of the model used in \cite{Chechkin2017} will be called ``the minimal model'' of diffusing diffusivity in what follows. Another model describing BnG diffusion is the diffusivity landscape model (DLM) \cite{Postnikov2020} which considers that the diffusion coefficient varies slowly in space. 

The possible connection between the diffusing diffusivity and DLM was stated in Ref.~\cite{Chechkin2017}: the temporal randomness of the diffusion coefficient can be considered as stemming from its spatial change along the trajectory of a diffusing particle, so that the ``minimal model'' is a kind of a mean-field approximation for the case of spatial changes. Even if the DD and DL models are gauged in such a way that they reproduce the main features of the phenomenon, their predictions differ in some details. Looking particularly into these details may deliver valuable experimental insights into the kind of disorder involved. Thus, the DLM (and other models with correlated spatial disorder like the one discussed in \cite{Luo2018}, not necessarily exhibiting the BnG behavior) show a pronounced central peak at the mode of the PDF of particles' displacements. This central peak is, however, absent in the minimal model. The existence of this central peak in \cite{Luo2018} was immediately connected to the correlated nature of disorder. 

Recently, in Ref.~\cite{Pacheco2021}, we concentrated on the behavior of the PDF of displacements close to its mode and showed that the PDF of displacements in several classical strongly disordered systems displays such a peak at its center. The behavior of this central peak is quite peculiar, since its presence shows that the convergence to a Gaussian  (i.e. normal) behavior under homogenization may follow a different pathway than the one commonly known from the Central Limit Theorem (CLT) applied to sums of many independent, identically distributed (i.i.d.) random variables following some continuous distribution (in our case this should be the short-time Laplace one). This standard situation suggests that the initially sharp peak would smoothen and lower. However, under homogenization, the central peak in the considered classical strongly disordered systems gets narrower under the rescaling $\mathbf{r} \to \mathbf{r}/\sqrt{t}$, $p \to t^{d/2} p$ implied by the CLT, while approximately keeping the height. Passing from the spatially disordered systems to their mean-filed counterparts (like the corresponding CTRWs, or the minimal model) restores the standard convergence pathway like the one predicted by the CLT. 

The differences in the convergence pathways have to do with the fact that some important local information about the system is erased when passing to the pre-averaged (mean-field) description. Now, one could ask, what is the important information erased? In the present work, we try to answer this question by simulating the particles' trajectories in DLM (described as a continuous-time random walk of particles on a lattice with position-dependent waiting times) and erase the correlations between the waiting times and positions, while fully preserving the temporal structure of the walk. The result of the discussion shows that the existence of the persistent peak is connected to spatiotemporal correlations, and destroying them (while fully preserving the temporal structure of the problem) leads to a different kind of behavior.  
We note that the answer to this question may apply in other similar situations in strongly disordered systems. 

The article is structured as follows: In Section~\ref{sec:DLM}, we revisit the diffusivity landscape model being the base of our investigation. Section~\ref{sec:space-time} explores the idea that the DLM presents strong spario-temporal correlations which ultimately leads to the PDF exhibiting a central peak. We show that destroying these spatiotemporal correlations while fully preserving the temporal structure of steps reproduces the PDF in DLM at short times, but leads to lowering and disappearing of the peak at long ones. Section~\ref{sec:corrCTRW} provides a CTRW model with correlated waiting times which partially reproduces the behavior found in this decorrelated DLM model, but fails to fully describe the situation. In Section~\ref{sec:checkerboard}, we discuss the role of the particular shape of the correlation function of diffusivities assumed in DLM by considering a slightly different model. Finally, Section~\ref{sec:conclu} presents concluding remarks. 
\color{black}

%---------------------------------------------------------------
%Section: Diffusivity landscape model
%---------------------------------------------------------------

\section{Diffusivity landscape model \label{sec:DLM}}

In what follows, we use the model proposed by Postnikov et al. \cite{Postnikov2020} which assumes the particles' diffusion in a heterogeneous medium modeled by a correlated diffusivity landscape $D(\mathbf{r})$.
This motion is described by the force-free Langevin equation with multiplicative noise
\begin{equation}
\frac{d}{dt} \mathbf{r} = \sqrt{2 D(\mathbf{r})} \; \pmb{\xi}(t),
\label{eq:langevin}
\end{equation}
with $\pmb{\xi}(t)$ being a Gaussian white noise with $\langle\pmb{\xi}(t)\rangle = 0$ and $\langle\xi_{\mu}(t)\xi_{\nu}(t^{\prime}) \rangle = \delta_{\mu \nu} \delta(t-t^{\prime})$ with $\mu, \nu$ representing Cartesian coordinates. 
This Langevin equation corresponds to the Fokker-Planck equation
\begin{equation}
\frac{\partial}{\partial t} p(\mathbf{r},t) = \nabla[ (1-\alpha) \nabla D(\mathbf{r}) + D(\mathbf{r}) \nabla ] p(\mathbf{r},t)
\label{eq:fokker_planck}
\end{equation}
with $\alpha$ being the interpretation parameter taking values in the interval ${0 \leq \alpha \leq 1}$ (see e.g. \cite{Arnoulx2023} for a comprehensive discussion). The authors of \cite{Postnikov2020} asked, under which condition would Eq.~(\ref{eq:fokker_planck}) describe the BnG diffusion, and found out that the two following conditions should be met: First, Eq~(\ref{eq:langevin}) without external potential must be interpreted in the Ito sense ($\alpha = 0$; then, of course, any other interpretation can be used by introducing the corresponding deterministic force \cite{Arnoulx2023}), and second, initial positions of diffusing particles must be sampled from the equilibrium distribution. Taking as a ``stylized fact'' that the PDF at short times has been observed to follow a Laplace distribution \cite{Wang2009,Leptos2009,Kurtuldu2011,Wang2012,Skaug2013,Yu2013,He2013,Guan2014,Thorneywork2016,He2016,Acharya2017,Wagner2017,Chakraborty2019,Kwon2019,Chakraborty2020,Pastore2021,Pastore2022}, one can then show that the single-point PDF of the diffusion coefficients in the corresponding landscape should be given by a Gamma distribution:
\begin{equation}
p(D) = \frac{\beta^{\beta}}{\Gamma(\beta)} \frac{1}{\overline{D}} \left( \frac{D}{\overline{D}}\right)^{\beta-1} \exp \left( - \beta \frac{D}{\overline{D}} \right),
\label{eq:dlm_pD}
\end{equation}
where $\Gamma(\cdot)$ is a Gamma function, and $\beta$ and $\overline{D}$ are shape parameters dependent on the dimension of space. 

In what follows, we will concentrate on the two-dimensional situation, for which $\beta = 5/2$ and $\overline{D} = 5 D_0 / 3$, with $D_0$ being the sampled diffusion coefficient, i.e., the one defining the slope of the ``experimental'' MSD assumed to strictly follow the linear dependence $\langle \mathbf{r}(t)^2 \rangle = 2 d D_0 t$ \cite{Postnikov2020}. 

A finite-difference discretization of the Fokker-Planck equation, Eq.~(\ref{eq:fokker_planck}), with $\alpha = 0$ on a square lattice with lattice constant $a$ leads to a master equation (see Eq.~(\ref{eq:master}) below) which, in its turn, defines a random walk scheme. The corresponding random walks are exactly what will be simulated in what follows. 

For $\alpha = 0$, Eq.~(\ref{eq:fokker_planck}) can be rewritten in the form 
\[
\frac{\partial}{\partial t} p(\mathbf{r},t) = \Delta [ D(\mathbf{r})  p(\mathbf{r},t) ],  
\]
and its discrete version is 
\begin{equation}
\frac{d}{dt} p_i(t) = \sum_{k=1}^4 \frac{D_{j_k}}{a^2} p_{j_k}(t) - \frac{4 D_i}{a^2} p_i(t).
\label{eq:master}
\end{equation}
Here, the discretization point $i$ corresponds to coordinates $(x_i,y_i)$ on a rectangular grid with the lattice constant $a$, and points $j_k$ are the four nearest neighbors of the lattice point $i$. 

Under the above discretization, the random diffusivity field at each lattice point translates into correlated values of local parameters $ D_i \equiv D(x_i,y_i)$, which are generated according to the following algorithm, Ref.~\cite{Postnikov2020}: One begins by constructing an array of independent Gaussian random variables $G_i$ with zero mean and unit variance. Then one generates a correlated Gaussian field $\widehat{G}_i$ by applying the Fourier filtering method \cite{Toral2014} to $G_i$. Like in \cite{Postnikov2020}, we take the correlation function of the correlated field to follow 
\begin{equation}
\rho(\mathbf{r}_{ij}) = \langle \widehat{G}_i \widehat{G}_j \rangle = \exp \left( - \frac{ \mathbf{r}_{ij}^2}{2\lambda^2} \right),
\label{eq:gauss_corr}
\end{equation}
with $\lambda$ being the correlation length, and $\mathbf{r}_{ij}$ the Euclidean distance between lattice points $i$ and $j$. We note that the choice of Eq.~(\ref{eq:gauss_corr}) is not dictated by any physical reasons but by the ease of numerical implementation and further calculations. In Section~\ref{sec:checkerboard}, we will explore the consequences of changing the correlation function of the diffusivity landscape by considering a checkerboard-like diffusivity landscape.

Finally, the correlated Gaussian field $\widehat{G}_i$ is transformed into the $\Gamma$-distributed diffusivity landscape $D_i$ by performing a probability transformation:
\begin{equation}
D_i = f(\widehat{G}) = F^{-1}_{\beta} \left\{ \frac{1}{2} \left[ 1 - \text{erf} \left( \frac{\widehat{G}}{\sqrt{2}} \right) \right] \right\},
\label{eq:change_var}
\end{equation}
where $\text{erf}(\cdot)$ is the error function and $F^{-1}_{\beta}(x)$ is the inverse of the cumulative distribution function (CDF) $F_{\beta}(D)$ for the PDF given by Eq.~(\ref{eq:dlm_pD}), which is given by
\[
F_{\beta}(D) = \int_0^D p(D') dD' = \frac{1}{\Gamma(\beta)} \gamma \left( \beta, \beta \frac{D}{\overline{D}} \right),
\]
with $\gamma(\cdot,\cdot)$ being the lower incomplete Gamma function. The procedure above generates a diffusivity landscape $D_i$ whose correlation function follows from that of the correlated Gaussian field, Eq.~(\ref{eq:gauss_corr}), by a transformation which will be discussed in Sec. \ref{sec:Corr}.

\begin{figure}[h!]
    \centering
    \def\svgwidth{0.8\columnwidth}
\begingroup%
  \makeatletter%
  \providecommand\color[2][]{%
    \errmessage{(Inkscape) Color is used for the text in Inkscape, but the package 'color.sty' is not loaded}%
    \renewcommand\color[2][]{}%
  }%
  \providecommand\transparent[1]{%
    \errmessage{(Inkscape) Transparency is used (non-zero) for the text in Inkscape, but the package 'transparent.sty' is not loaded}%
    \renewcommand\transparent[1]{}%
  }%
  \providecommand\rotatebox[2]{#2}%
  \newcommand*\fsize{\dimexpr\f@size pt\relax}%
  \newcommand*\lineheight[1]{\fontsize{\fsize}{#1\fsize}\selectfont}%
  \ifx\svgwidth\undefined%
    \setlength{\unitlength}{460.79998848bp}%
    \ifx\svgscale\undefined%
      \relax%
    \else%
      \setlength{\unitlength}{\unitlength * \real{\svgscale}}%
    \fi%
  \else%
    \setlength{\unitlength}{\svgwidth}%
  \fi%
  \global\let\svgwidth\undefined%
  \global\let\svgscale\undefined%
  \makeatother%
  \begin{picture}(1,0.75)%
    \lineheight{1}%
    \setlength\tabcolsep{0pt}%
    \put(0,0){\includegraphics[width=\unitlength,page=1]{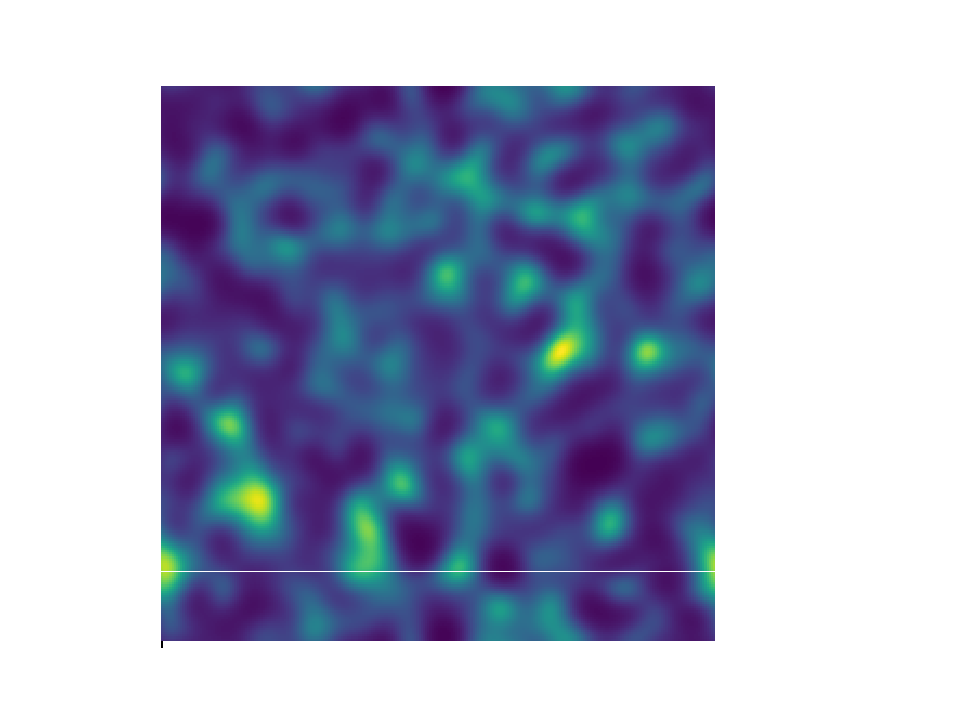}}%
    \put(0.16172786,0.04082964){\color[rgb]{0,0,0}\makebox(0,0)[lt]{\lineheight{1.25}\smash{\begin{tabular}[t]{l}\small $0$\end{tabular}}}}%
    \put(0,0){\includegraphics[width=\unitlength,page=2]{landscape.pdf}}%
    \put(0.37351345,0.04082964){\color[rgb]{0,0,0}\makebox(0,0)[lt]{\lineheight{1.25}\smash{\begin{tabular}[t]{l}\small $100$\end{tabular}}}}%
    \put(0.48992824,0.04955162){\color[rgb]{0,0,0}\makebox(0,0)[lt]{\lineheight{1.25}\smash{\begin{tabular}[t]{l}$x$\end{tabular}}}}%
    \put(0,0){\includegraphics[width=\unitlength,page=3]{landscape.pdf}}%
    \put(0.59909939,0.04082964){\color[rgb]{0,0,0}\makebox(0,0)[lt]{\lineheight{1.25}\smash{\begin{tabular}[t]{l}\small $200$\end{tabular}}}}%
    \put(0,0){\includegraphics[width=\unitlength,page=4]{landscape.pdf}}%
    \put(0.13850825,0.65288628){\color[rgb]{0,0,0}\makebox(0,0)[lt]{\lineheight{1.25}\smash{\begin{tabular}[t]{l}\small $0$\end{tabular}}}}%
    \put(0,0){\includegraphics[width=\unitlength,page=5]{landscape.pdf}}%
    \put(0.1009069,0.42730035){\color[rgb]{0,0,0}\makebox(0,0)[lt]{\lineheight{1.25}\smash{\begin{tabular}[t]{l}\small $100$\end{tabular}}}}%
    \put(0.08898588,0.32652778){\color[rgb]{0,0,0}\rotatebox{0}{\makebox(0,0)[lt]{\lineheight{1.25}\smash{\begin{tabular}[t]{l}$y$\end{tabular}}}}}%
    \put(0,0){\includegraphics[width=\unitlength,page=6]{landscape.pdf}}%
    \put(0.1009069,0.20171528){\color[rgb]{0,0,0}\makebox(0,0)[lt]{\lineheight{1.25}\smash{\begin{tabular}[t]{l}\small $200$\end{tabular}}}}%
    \put(0,0){\includegraphics[width=\unitlength,page=7]{landscape.pdf}}%
    \put(0.82781684,0.15105447){\color[rgb]{0,0,0}\makebox(0,0)[lt]{\lineheight{1.25}\smash{\begin{tabular}[t]{l}$1$\end{tabular}}}}%
    \put(0,0){\includegraphics[width=\unitlength,page=8]{landscape.pdf}}%
    \put(0.82781684,0.23472873){\color[rgb]{0,0,0}\makebox(0,0)[lt]{\lineheight{1.25}\smash{\begin{tabular}[t]{l}$2$\end{tabular}}}}%
    \put(0,0){\includegraphics[width=\unitlength,page=9]{landscape.pdf}}%
    \put(0.82781684,0.31840278){\color[rgb]{0,0,0}\makebox(0,0)[lt]{\lineheight{1.25}\smash{\begin{tabular}[t]{l}$3$\end{tabular}}}}%
    \put(0,0){\includegraphics[width=\unitlength,page=10]{landscape.pdf}}%
    \put(0.82781684,0.40207899){\color[rgb]{0,0,0}\makebox(0,0)[lt]{\lineheight{1.25}\smash{\begin{tabular}[t]{l}$4$\end{tabular}}}}%
    \put(0,0){\includegraphics[width=\unitlength,page=11]{landscape.pdf}}%
    \put(0.82781684,0.48575304){\color[rgb]{0,0,0}\makebox(0,0)[lt]{\lineheight{1.25}\smash{\begin{tabular}[t]{l}$5$\end{tabular}}}}%
    \put(0,0){\includegraphics[width=\unitlength,page=12]{landscape.pdf}}%
    \put(0.82781684,0.56942708){\color[rgb]{0,0,0}\makebox(0,0)[lt]{\lineheight{1.25}\smash{\begin{tabular}[t]{l}$6$\end{tabular}}}}%
    \put(0.83667824,0.65564352){\color[rgb]{0,0,0}\rotatebox{0}{\makebox(0,0)[lt]{\lineheight{1.25}\smash{\begin{tabular}[t]{l}\large $D(x,y)$\end{tabular}}}}}%
  \end{picture}%
\endgroup%
    \caption{A two-dimensional realization of the diffusivity landscape $D(\mathbf{r})$ in the diffusivity landscape model. It corresponds to a $256 \times 256$ lattice with correlation length $\lambda = 10$ and sampled diffusion coefficient $D_0 = 1$.}
    \label{fig:landscape}
\end{figure}

Figure~\ref{fig:landscape} shows a realization of the diffusivity landscape $D_i$ for a lattice of $256 \times 256$ with $a = 1$, $D_0 = 1$ and $\lambda = 10$.

Let us now return to our Eq.~(\ref{eq:master}). Defining the transition rates as
\[
\omega_{i \to j} = \frac{D_i}{a^2} = \frac{1}{4} \left(\frac{a^2}{4 D_i} \right)^{-1} = \frac{1}{4} \frac{1}{\tau_i},
\]
with
\begin{equation}
\tau_i = \frac{a^2}{4 D_i}
\label{eq:tau}
\end{equation}
being the mean waiting time at a site, and $1/4$ corresponding to the probability to choose one of the four neighbors to jump to. We put 
Eq.~(\ref{eq:master}) into a standard form of a master equation
\[
 \frac{d}{dt} p_i(t) = \sum_{k=1}^4 \omega_{j_k \to i}  p_{j_k}(t) - \sum_{k=1}^4 \omega_{i \to j_k}  p_{i}(t),
\]
which can be rewritten as
\begin{equation}
\frac{d}{dt} p_i(t) = \frac{1}{4} \sum_{k=1}^4  \frac{1}{\tau_{j_k}}  p_{j_k}(t) - \frac{1}{\tau_i} p_i(t).
\label{eq:master_1}
\end{equation}
In Ref.~\cite{Postnikov2020}, Eq.~(\ref{eq:master_1}) was solved using the forward Euler method. Here, we employ another approach. Like in Ref.~\cite{Pacheco2021}, we use the fact that the master equation~(\ref{eq:master_1}) corresponds to a CTRW with exponential waiting times distribution \cite{Klafter2011}. Thus, to solve the master equation, i.e., to obtain the evolution of the PDF, we generate random walk trajectories whose waiting times follow the exponential waiting time density
\[
\psi(t \arrowvert \tau_i) = \frac{1}{\tau_i} \exp \left( - \frac{t}{\tau_i} \right),
\]
with $\tau_i$ given by Eq.~(\ref{eq:tau}). Taking the lattice spacing to be the length unit of the problem ($a = 1$), we get
\begin{equation}
\psi(t\arrowvert D_i) = 4 D_i \exp \left( - 4 D_i t \right).
\label{eq:waiting_times_site}
\end{equation}
Note that the single-step displacements in our CTRW are i.i.d. random variables (each step has a unit length and arbitrary, random direction), while waiting times are not independent since $D_i$ at neighboring points are correlated. An illustration of the procedure to generate random trajectories can be seen in panel $(a)$ of Figure~\ref{fig:trajectory}. As we shall see, this alternative method allows us to study the role of space-time correlations in the DLM, which would be impossible to do by solving the ordinary differential equations (ODEs). Moreover, generating random trajectories is considerably less computationally expensive than solving ODEs, and allows us to have much better statistics of the desired quantities.

%---------------------------------------------------------------
%Section: Space-time correlations
%---------------------------------------------------------------

\section{Space-time correlations \label{sec:space-time}}

In the last decades, several correlated CTRW models were proposed which lead to interesting behaviors \cite{Chechkin2009,Tejedor2010,Schulz2013}. However, all these correlated models focus only on either the temporal part or the spatial part separately or simultaneously, but leave the spatiotemporal correlations out of the picture. The reason for this is that dealing with such cross-correlations is, in general, a very complex task, even from a computational point of view. 

All (semi-)analytical results usually come from applying mean-field techniques which partially or completely ignore the fine-scale structure of the system, so that some interesting features of the spatially disordered systems are not reproduced. This is the case, e.g., for the behavior of the central peak seen in the DLM \cite{Pacheco2021}, which is not reproduced in such pre-averaged models like the CTRW description of the DLM \cite{Pacheco2021}, or the minimal model of BnG \cite{Chechkin2017}. It is in this regard that we seek to know to what extent the spatiotemporal correlations are responsible for the persistence of the central peak and the unusual art of convergence to a Gaussian distribution by narrowing of the central peak under rescaling $\mathbf{r} \to \mathbf{r} / \sqrt{t}$, $p \to t^{d/2} p$, instead of its lowering. To assess the effects of spatiotemporal correlations in the DLM, we remove them by randomization of step directions and see what changes by comparing the PDF in the decorrelated motion with that in the correlated one. 

\begin{figure}[h!]
    \centering
    \def\svgwidth{1.0\columnwidth}
\begingroup%
  \makeatletter%
  \providecommand\color[2][]{%
    \errmessage{(Inkscape) Color is used for the text in Inkscape, but the package 'color.sty' is not loaded}%
    \renewcommand\color[2][]{}%
  }%
  \providecommand\transparent[1]{%
    \errmessage{(Inkscape) Transparency is used (non-zero) for the text in Inkscape, but the package 'transparent.sty' is not loaded}%
    \renewcommand\transparent[1]{}%
  }%
  \providecommand\rotatebox[2]{#2}%
  \newcommand*\fsize{\dimexpr\f@size pt\relax}%
  \newcommand*\lineheight[1]{\fontsize{\fsize}{#1\fsize}\selectfont}%
  \ifx\svgwidth\undefined%
    \setlength{\unitlength}{166.49380346bp}%
    \ifx\svgscale\undefined%
      \relax%
    \else%
      \setlength{\unitlength}{\unitlength * \real{\svgscale}}%
    \fi%
  \else%
    \setlength{\unitlength}{\svgwidth}%
  \fi%
  \global\let\svgwidth\undefined%
  \global\let\svgscale\undefined%
  \makeatother%
  \begin{picture}(1,0.43263291)%
    \lineheight{1}%
    \setlength\tabcolsep{0pt}%
    \put(0,0){\includegraphics[width=\unitlength,page=1]{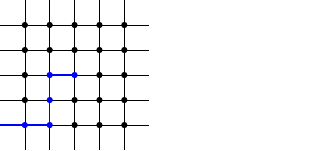}}%
    \put(0.0,0.45){\color[rgb]{0,0,0}\makebox(0,0)[lt]{\lineheight{1.25}\smash{\begin{tabular}[t]{l}$(a)$\end{tabular}}}}%
    \put(0.96,0.45){\color[rgb]{0,0,0}\makebox(0,0)[lt]{\lineheight{1.25}\smash{\begin{tabular}[t]{l}$(b)$\end{tabular}}}}%
    \put(0.24053004,0.24930659){\color[rgb]{0,0,1}\makebox(0,0)[lt]{\lineheight{1.25}\smash{\begin{tabular}[t]{l}\scriptsize $\mathbf{r}_0,t_0$\end{tabular}}}}% %a
    \put(0,0){\includegraphics[width=\unitlength,page=2]{trajectory.pdf}}%
    \put(0.70024302,0.24930659){\color[rgb]{1,0,0}\makebox(0,0)[lt]{\lineheight{1.25}\smash{\begin{tabular}[t]{l}\scriptsize $\mathbf{r}_0,t_0$\end{tabular}}}}% %b
    \put(0.08702698,0.24930659){\color[rgb]{0,0,1}\makebox(0,0)[lt]{\lineheight{1.25}\smash{\begin{tabular}[t]{l}\scriptsize $\mathbf{r}_1,t_1$\end{tabular}}}}%
    \put(0.16190748,0.17056081){\color[rgb]{0,0,1}\makebox(0,0)[lt]{\lineheight{1.25}\smash{\begin{tabular}[t]{l}\scriptsize $\mathbf{r}_2,t_2$\end{tabular}}}}%
    \put(0.16190748,0.09559859){\color[rgb]{0,0,1}\makebox(0,0)[lt]{\lineheight{1.25}\smash{\begin{tabular}[t]{l}\scriptsize $\mathbf{r}_3,t_3$\end{tabular}}}}%
    \put(0.00702698,0.09559859){\color[rgb]{0,0,1}\makebox(0,0)[lt]{\lineheight{1.25}\smash{\begin{tabular}[t]{l}\scriptsize $\mathbf{r}_4,t_4$\end{tabular}}}}%
    \put(0.85560556,0.24930659){\color[rgb]{1,0,0}\makebox(0,0)[lt]{\lineheight{1.25}\smash{\begin{tabular}[t]{l}\scriptsize $\mathbf{r}_1^{\prime},t_1$\end{tabular}}}}%
    \put(0.85333033,0.33046756){\color[rgb]{1,0,0}\makebox(0,0)[lt]{\lineheight{1.25}\smash{\begin{tabular}[t]{l}\scriptsize $\mathbf{r}_2^{\prime},t_2$\end{tabular}}}}%
    \put(0.70024302,0.33046756){\color[rgb]{1,0,0}\makebox(0,0)[lt]{\lineheight{1.25}\smash{\begin{tabular}[t]{l}\scriptsize $\mathbf{r}_3^{\prime},t_3$\end{tabular}}}}%
    \put(0.70024302,0.4138611){\color[rgb]{1,0,0}\makebox(0,0)[lt]{\lineheight{1.25}\smash{\begin{tabular}[t]{l}\scriptsize $\mathbf{r}_4^{\prime},t_4$\end{tabular}}}}%
  \end{picture}%
\endgroup%
    \caption{Schematics of the procedure to decouple space and time. The real particle follows the blue trajectory, whereas the decoupled particle follows the red trajectory. Notice that the waiting times for both particles are the same, and they depend only on the positions of the real particle.}
    \label{fig:trajectory}
\end{figure}

Let us consider a particle whose initial position is $\mathbf{r}_0$, and follow its true motion as given by a random walk scheme corresponding to the master equation (\ref{eq:master}). At that position, the particle waits for a time $t_0$ which is drawn from the exponential distribution  $\psi(t\arrowvert D_0)$ given in Eq.~(\ref{eq:waiting_times_site}), with $D_0 = D(\mathbf{r}_0)$. Next, the particle randomly jumps to one of its neighboring lattice points whose position is $\mathbf{r}_1$, and then waits for another time $t_1$ which is now drawn from the exponential distributions $\psi(t\arrowvert D_1)$, with $D_1 = D(\mathbf{r}_1)$. This process of jumping and waiting is repeated until the maximal simulation time $t_{max}$ is exceeded by the sum of waiting times. At the end, the trajectory of our particle (which we will call \textit{real} particle to distinguish its motion from its randomized counterparts) is given by a list of positions $\{ \mathbf{r}_0, \mathbf{r}_1, \mathbf{r}_2, \dots \}$,
which correspond to a simple random walk, and a list $\{ t_0, t_1,t_2, \dots \}$ of the corresponding waiting times between subsequent jumps which are drawn from the exponential distributions $\psi(t\arrowvert D_{i})$ with $D_i = D(\mathbf{r}_i)$, and 
which are therefore dependent on the particles' positions. This dependence of the exponential distribution on the value of the local diffusion coefficient generates the spatiotemporal correlations in the DLM. The procedure to obtain the trajectories of a real particle is sketched in panel $(a)$ of Figure~\ref{fig:trajectory}. 

We now use the above trajectories of true motion of particles, which we refer to as ``real trajectories'' in what follows, to generate new trajectories in which space and time are uncorrelated. Let us start by taking the temporal part of a real trajectory, i.e., the list of waiting times $\{ t_0, t_1, \dots \}$, and discard the spatial part. Then we proceed as follows: Let us consider a new particle, which we will call the \textit{decoupled} particle, whose initial position is the same as for the real one, i.e., $\mathbf{r}_0$. The decoupled particle then waits a time equal to the first waiting time of the real particle, namely $t_0$, and makes a jump to one of the neighboring lattice points with position $\mathbf{r}^{\prime}_1$. At this position, the decoupled particle waits a time equal to the second waiting time of the real particle, namely $t_1$, to make the next jump in a random direction. This process is repeated until the same number of steps as for a real particle is done, and the maximal time $t_{max}$ is exceeded. Hence, one ends up with a trajectory for the decoupled particle consisting of a list of the positions $\{ \mathbf{r}_0, \mathbf{r}^{\prime}_1, \mathbf{r}^{\prime}_2, \dots \}$ being sums of i.i.d. random steps, and the same list of waiting times $\{ t_0, t_1, \dots \}$ as for the real one, which are however decoupled from the corresponding particle's positions.  This process is depicted in panel $(b)$ of Figure~\ref{fig:trajectory}. 
The trajectories of real and decoupled particles are then used for obtaining the PDFs of displacements in a given realization of the landscape. Similar PDFs are obtained for different realizations of the diffusivity landscapes and then weighted-averaged under the equilibrium condition: the corresponding weight is proportional to the waiting time $t_0$ at $\mathbf{r}_0$ in the corresponding landscape.

\begin{figure}[h!]
    \centering
    \def\svgwidth{\columnwidth}
\begingroup%
  \makeatletter%
  \providecommand\color[2][]{%
    \errmessage{(Inkscape) Color is used for the text in Inkscape, but the package 'color.sty' is not loaded}%
    \renewcommand\color[2][]{}%
  }%
  \providecommand\transparent[1]{%
    \errmessage{(Inkscape) Transparency is used (non-zero) for the text in Inkscape, but the package 'transparent.sty' is not loaded}%
    \renewcommand\transparent[1]{}%
  }%
  \providecommand\rotatebox[2]{#2}%
  \newcommand*\fsize{\dimexpr\f@size pt\relax}%
  \newcommand*\lineheight[1]{\fontsize{\fsize}{#1\fsize}\selectfont}%
  \ifx\svgwidth\undefined%
    \setlength{\unitlength}{1440bp}%
    \ifx\svgscale\undefined%
      \relax%
    \else%
      \setlength{\unitlength}{\unitlength * \real{\svgscale}}%
    \fi%
  \else%
    \setlength{\unitlength}{\svgwidth}%
  \fi%
  \global\let\svgwidth\undefined%
  \global\let\svgscale\undefined%
  \makeatother%
  \begin{picture}(1,0.58333333)%
    \lineheight{1}%
    \setlength\tabcolsep{0pt}%
    \put(0,0){\includegraphics[width=\unitlength,page=1]{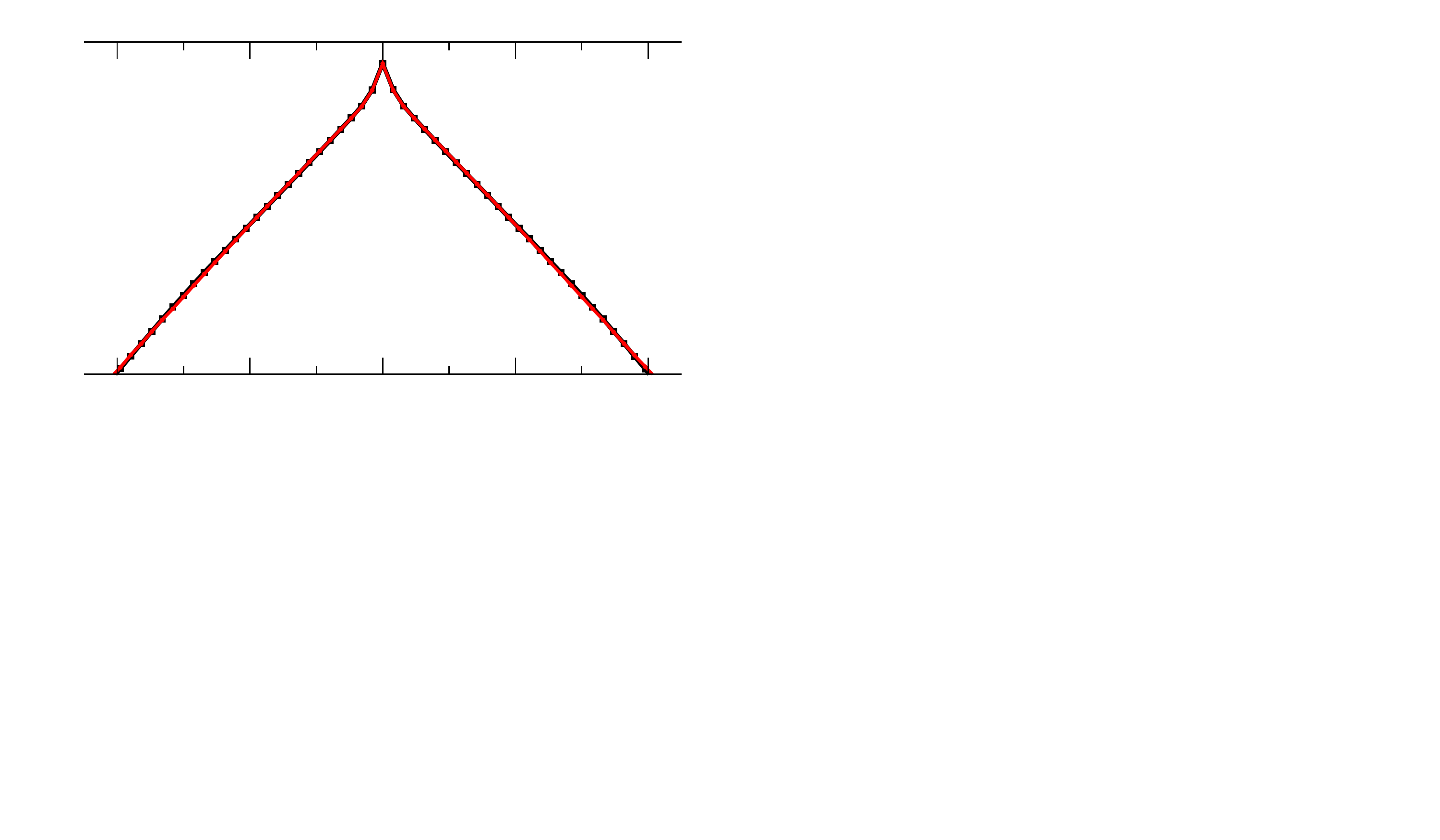}}%
    \put(0.0650475,0.30648486){\makebox(0,0)[lt]{\lineheight{1.25}\smash{\begin{tabular}[t]{l}\tiny $-8$\end{tabular}}}}%
    \put(0.1569875,0.30648486){\makebox(0,0)[lt]{\lineheight{1.25}\smash{\begin{tabular}[t]{l}\tiny $-4$\end{tabular}}}}%
    \put(0.26184417,0.30648486){\makebox(0,0)[lt]{\lineheight{1.25}\smash{\begin{tabular}[t]{l}\tiny $0$\end{tabular}}}}%
    \put(0.3540341,0.30648486){\makebox(0,0)[lt]{\lineheight{1.25}\smash{\begin{tabular}[t]{l}\tiny $4$\end{tabular}}}}%
    \put(0.4464741,0.30648486){\makebox(0,0)[lt]{\lineheight{1.25}\smash{\begin{tabular}[t]{l}\tiny $8$\end{tabular}}}}%
    \put(0.30184417,0.29298486){\makebox(0,0)[lt]{\lineheight{1.25}\smash{\begin{tabular}[t]{l}\scriptsize $\xi$\end{tabular}}}}%
    \put(0,0){\includegraphics[width=\unitlength,page=2]{fake_real.pdf}}%
    \put(0.00816667,0.36420451){\makebox(0,0)[lt]{\lineheight{1.25}\smash{\begin{tabular}[t]{l}\tiny $10^{-4}$\end{tabular}}}}%
    \put(0.00816667,0.45647729){\makebox(0,0)[lt]{\lineheight{1.25}\smash{\begin{tabular}[t]{l}\tiny $10^{-2}$\end{tabular}}}}%
    \put(0.02816667,0.54858333){\makebox(0,0)[lt]{\lineheight{1.25}\smash{\begin{tabular}[t]{l}\tiny $10^{0}$\end{tabular}}}}%
    \put(0.00291667,0.41557576){\rotatebox{0}{\makebox(0,0)[lt]{\lineheight{1.25}\smash{\begin{tabular}[t]{l}\scriptsize $q(\xi)$\end{tabular}}}}}%
    \put(0,0){\includegraphics[width=\unitlength,page=3]{fake_real.pdf}}%
    \put(0.34684417,0.51616667){\makebox(0,0)[lt]{\lineheight{1.25}\smash{\begin{tabular}[t]{l}\tiny $(a) \; t = 10^1$\end{tabular}}}}%
    \put(0,0){\includegraphics[width=\unitlength,page=4]{fake_real.pdf}}%
    \put(0.5628734,0.30648486){\makebox(0,0)[lt]{\lineheight{1.25}\smash{\begin{tabular}[t]{l}\tiny $-8$\end{tabular}}}}%
    \put(0.6548134,0.30648486){\makebox(0,0)[lt]{\lineheight{1.25}\smash{\begin{tabular}[t]{l}\tiny $-4$\end{tabular}}}}%
    \put(0.75967007,0.30648486){\makebox(0,0)[lt]{\lineheight{1.25}\smash{\begin{tabular}[t]{l}\tiny $0$\end{tabular}}}}%
    \put(0.85186007,0.30648486){\makebox(0,0)[lt]{\lineheight{1.25}\smash{\begin{tabular}[t]{l}\tiny $4$\end{tabular}}}}%
    \put(0.9443,0.30648486){\makebox(0,0)[lt]{\lineheight{1.25}\smash{\begin{tabular}[t]{l}\tiny $8$\end{tabular}}}}%
    \put(0.79967007,0.29298486){\makebox(0,0)[lt]{\lineheight{1.25}\smash{\begin{tabular}[t]{l}\scriptsize $\xi$\end{tabular}}}}%
    \put(0,0){\includegraphics[width=\unitlength,page=5]{fake_real.pdf}}%
    \put(0.50599257,0.36420451){\makebox(0,0)[lt]{\lineheight{1.25}\smash{\begin{tabular}[t]{l}\tiny $10^{-4}$\end{tabular}}}}%
    \put(0.50599257,0.45647729){\makebox(0,0)[lt]{\lineheight{1.25}\smash{\begin{tabular}[t]{l}\tiny $10^{-2}$\end{tabular}}}}%
    \put(0.5238259,0.54858333){\makebox(0,0)[lt]{\lineheight{1.25}\smash{\begin{tabular}[t]{l}\tiny $10^{0}$\end{tabular}}}}%
    \put(0.50074257,0.41557576){\rotatebox{0}{\makebox(0,0)[lt]{\lineheight{1.25}\smash{\begin{tabular}[t]{l}\scriptsize $q(\xi)$\end{tabular}}}}}%
    \put(0,0){\includegraphics[width=\unitlength,page=6]{fake_real.pdf}}%
    \put(0.84467007,0.51616667){\makebox(0,0)[lt]{\lineheight{1.25}\smash{\begin{tabular}[t]{l}\tiny $(b) \; t = 10^2$\end{tabular}}}}%
    \put(0,0){\includegraphics[width=\unitlength,page=7]{fake_real.pdf}}%
    \put(0.0650475,0.02966667){\makebox(0,0)[lt]{\lineheight{1.25}\smash{\begin{tabular}[t]{l}\tiny $-8$\end{tabular}}}}%
    \put(0.1569875,0.02966667){\makebox(0,0)[lt]{\lineheight{1.25}\smash{\begin{tabular}[t]{l}\tiny $-4$\end{tabular}}}}%
    \put(0.26184417,0.02966667){\makebox(0,0)[lt]{\lineheight{1.25}\smash{\begin{tabular}[t]{l}\tiny $0$\end{tabular}}}}%
    \put(0.3540341,0.02966667){\makebox(0,0)[lt]{\lineheight{1.25}\smash{\begin{tabular}[t]{l}\tiny $4$\end{tabular}}}}%
    \put(0.4464741,0.02966667){\makebox(0,0)[lt]{\lineheight{1.25}\smash{\begin{tabular}[t]{l}\tiny $8$\end{tabular}}}}%
    \put(0.30184417,0.01616667){\makebox(0,0)[lt]{\lineheight{1.25}\smash{\begin{tabular}[t]{l}\scriptsize $\xi$\end{tabular}}}}%
    \put(0,0){\includegraphics[width=\unitlength,page=8]{fake_real.pdf}}%
    \put(0.00816667,0.08738639){\makebox(0,0)[lt]{\lineheight{1.25}\smash{\begin{tabular}[t]{l}\tiny $10^{-4}$\end{tabular}}}}%
    \put(0.00816667,0.1796591){\makebox(0,0)[lt]{\lineheight{1.25}\smash{\begin{tabular}[t]{l}\tiny $10^{-2}$\end{tabular}}}}%
    \put(0.02816667,0.27176514){\makebox(0,0)[lt]{\lineheight{1.25}\smash{\begin{tabular}[t]{l}\tiny $10^{0}$\end{tabular}}}}%
    \put(0.00291667,0.13875757){\rotatebox{0}{\makebox(0,0)[lt]{\lineheight{1.25}\smash{\begin{tabular}[t]{l}\scriptsize $q(\xi)$\end{tabular}}}}}%
    \put(0,0){\includegraphics[width=\unitlength,page=9]{fake_real.pdf}}%
    \put(0.34684417,0.23934847){\makebox(0,0)[lt]{\lineheight{1.25}\smash{\begin{tabular}[t]{l}\tiny $(c) \; t = 10^3$\end{tabular}}}}%
    \put(0,0){\includegraphics[width=\unitlength,page=10]{fake_real.pdf}}%
    \put(0.5628734,0.02966667){\makebox(0,0)[lt]{\lineheight{1.25}\smash{\begin{tabular}[t]{l}\tiny $-8$\end{tabular}}}}%
    \put(0.6548134,0.02966667){\makebox(0,0)[lt]{\lineheight{1.25}\smash{\begin{tabular}[t]{l}\tiny $-4$\end{tabular}}}}%
    \put(0.75967007,0.02966667){\makebox(0,0)[lt]{\lineheight{1.25}\smash{\begin{tabular}[t]{l}\tiny $0$\end{tabular}}}}%
    \put(0.85186007,0.02966667){\makebox(0,0)[lt]{\lineheight{1.25}\smash{\begin{tabular}[t]{l}\tiny $4$\end{tabular}}}}%
    \put(0.9443,0.02966667){\makebox(0,0)[lt]{\lineheight{1.25}\smash{\begin{tabular}[t]{l}\tiny $8$\end{tabular}}}}%
    \put(0.79967007,0.01616667){\makebox(0,0)[lt]{\lineheight{1.25}\smash{\begin{tabular}[t]{l}\scriptsize $\xi$\end{tabular}}}}%
    \put(0,0){\includegraphics[width=\unitlength,page=11]{fake_real.pdf}}%
    \put(0.50599257,0.08738639){\makebox(0,0)[lt]{\lineheight{1.25}\smash{\begin{tabular}[t]{l}\tiny $10^{-4}$\end{tabular}}}}%
    \put(0.50599257,0.1796591){\makebox(0,0)[lt]{\lineheight{1.25}\smash{\begin{tabular}[t]{l}\tiny $10^{-2}$\end{tabular}}}}%
    \put(0.5238259,0.27176514){\makebox(0,0)[lt]{\lineheight{1.25}\smash{\begin{tabular}[t]{l}\tiny $10^{0}$\end{tabular}}}}%
    \put(0.50074257,0.13875757){\rotatebox{0}{\makebox(0,0)[lt]{\lineheight{1.25}\smash{\begin{tabular}[t]{l}\scriptsize $q(\xi)$\end{tabular}}}}}%    
    \put(0,0){\includegraphics[width=\unitlength,page=12]{fake_real.pdf}}%
    \put(0.84467007,0.23934847){\makebox(0,0)[lt]{\lineheight{1.25}\smash{\begin{tabular}[t]{l}\tiny $(d) \; t = 10^4$\end{tabular}}}}%
  \end{picture}%
\endgroup%
    \caption{A comparison of the one-dimensional cut of the PDF $q(\xi) = p(x,0) t$ of rescaled displacements $\xi = x / \sqrt{t}$ for the real particle (black) and decoupled particle (red), see text for details. Each panel represents a particular time. The flattening of the the central peak in the PDF of positions of the decoupled particle is evident at longer times.} 
    \label{fig:fake_real}
\end{figure}

The resulting PDFs for real and decoupled particles are displayed in Figure~\ref{fig:fake_real}. The four panels of the plot present four different maximal times: $t_{max}=10^1$, $10^2$, $10^3$, and $10^4$. Each panel presents a comparison of the PDF for the real (black dots) and decoupled (red dots) particles. Each PDF is the average over $10^4$ realizations of the diffusivity landscape over a lattice of $2048 \times 2048$ with $\lambda = 10$ and $D_0 = 1$. Each realization contains $10^5$ particles. Plotted in Figure~\ref{fig:fake_real} is a cut of PDF $p(x,y;t)$ through the origin at $y = 0$. Moreover, following \cite{Pacheco2021}, we plot the PDF as a function of the rescaled displacement $\xi = x/\sqrt{t}$. To keep the normalization of the PDF, it has to be rescaled as $q(\xi) = t \cdot p(\xi)$. Figure~\ref{fig:fake_real} shows that the decoupling of the spatial and temporal aspects of the motion changes the art of convergence to the Gaussian from the unusual one, by narrowing of the central peak, to the CLT-like convergence, by lowering and smoothening the peak. 

%\color{blue}
In Refs. \cite{Luo2018} and \cite{Pacheco2021}, the existence of the central peak was connected with the set of particles which started their motion in a patch with a very low local diffusivity, so that they could hardly leave the patch 
until very long times. The randomization results show that this is only a partial explanation, since at the beginning of its motion a decoupled particle experinces the same, very long waiting times as the real one provided it 
started in such a patch. The trajectory of a decoupled particle is simply a different realization of a random walk with the same starting point associated with the same list of waiting times, so that only kind of correlations which are destroyed by our
procedure correspond to what happens if the particle returns to a close vicinity of its initial position after making an excursion to the outside of the patch (a real particle will again experience long waiting times, while for a decoupled one these 
new waiting times are not necessarily long, and new long waiting periods occure at different positions). Thus, it is a behaviour after an excursion that makes the peak persistent.
%\color{black}

Figure~\ref{fig:fake_real}, however, shows that the central peak for a decoupled motion is still present at a times as long as $10^3$. The reason for its existence may only be connected with correlations between waiting times 
along the trajectory, which are not destroyed by decoupling. Therefore, our next step will be to include the temporal correlations into a space-time-decoupled CTRW model.

%---------------------------------------------------------------
%Section: Correlated CTRW
%---------------------------------------------------------------

\section{Time-correlated continuous-time random walk \label{sec:corrCTRW}}

Ref.~\cite{Pacheco2021} presented a mean-field description of the DLM. This mean-field description is constructed as an \textit{uncorrelated} CTRW model whose waiting time distribution is found by averaging the mean waiting time distribution at a site (Eq.~(\ref{eq:waiting_times_site})) over the distribution of the diffusion coefficients, which is given by the one-point distribution of the diffusivity landscape (Eq.~(\ref{eq:dlm_pD})). This mean-field waiting time distribution is given by
\begin{equation}
\psi(t) = \int_0^{\infty} \psi(t\arrowvert D)  p(D) dD = \frac{5}{2} \left( \frac{3}{8} \right)^{5/2} \left( \frac{3}{8} + t \right)^{-7/2},
\label{eq:CTRW_wtd}
\end{equation}
for $D_0 = 1$. The corresponding waiting time density corresponds to a Pareto type II distribution with mean waiting time $\langle t \rangle = 1/4$ and second moment $\langle t^2 \rangle = 3/8$. The fact that the initial state of the system in the DLM must be at equilibrium should also be included in its mean-field description. This is done by taking the first waiting time to follow the PDF \cite{Klafter2011}
\begin{equation}
\psi_1(t) = \frac{1}{\langle t \rangle} \left[ 1 - \int_0^t \psi(t^{\prime}) d t^{\prime} \right] = \frac{3}{2} \left( \frac{3}{8} \right)^{3/2} \left( \frac{3}{8} + t \right)^{-5/2},
\label{eq:CTRW_wtd_1}
\end{equation}
which is also a Pareto type II distribution with a different exponent. 

As we have shown in \cite{Pacheco2021}, this mean-field description, being a pre-averaged model (cf. Eq.~(\ref{eq:CTRW_wtd})) neglecting \textit{all} correlations, does not show any peak at the center of the distribution, except for 
a decaying remains of the initial condition at $\mathbf{r}(0) = 0$. The PDF of the particles' displacements in this model is shown in Figure~\ref{fig:fake_ctrw} to be compared with the results for a decoupled particle and for a CTRW model
reproducing the serial correlations along the trajectory discussed below.

%---------------------------------------------------------------
%Subsection: Correlation along trajectories
%---------------------------------------------------------------

\subsection{Correlation function of the diffusion coefficient along the trajectories \label{sec:Corr}}

We would like to know to what extent the PDF for the decoupled particle can be replicated if temporal correlations are included. To do so, let us first determine the correlations of the diffusion coefficients along the trajectories of the random walk.

Let us start by finding an approximation for the correlation function of the diffusivity landscape $D(\mathbf{r})$ in terms of that of the correlated Gaussian field $\widehat{G}$ defined in Eq.~(\ref{eq:gauss_corr}). The correlation function of the diffusivity landscape is, by definition,
\begin{equation}
C_{DD}(\mathbf{r}) \equiv \xi(\mathbf{r}) =  \frac{\langle \delta D(\mathbf{0}) \delta D(\mathbf{r}) \rangle}{\sigma_D^2} = \frac{\langle D(\mathbf{0}) D(\mathbf{r}) \rangle - \overline{D}^2}{\sigma_D^2},
\label{eq:DL_corr}
\end{equation}
with  $\sigma_D^2$ the variance of the local diffusivity, and $\delta D(\mathbf{r}) = D(\mathbf{r}) - \overline{D}$. In our case, $\overline{D} = 5/3$ and $\sigma_D^2 = 10 / 9$, for $D_0 = 1$ in two dimensions. 

Let us turn our attention to the mean $\langle D(\mathbf{0}) D(\mathbf{r}) \rangle$ in the last expression of Eq.~(\ref{eq:DL_corr}). Just for convenience, let us denote $D(\mathbf{0})$ and $D(\mathbf{r})$ as $D_1$ and $D_2$, respectively. By doing so, one can write 
\begin{equation}
\langle D(\mathbf{0}) D(\mathbf{r}) \rangle = \langle D_1 D_2 \rangle  = \int_0^{\infty} \int_0^{\infty} dD_1 dD_2 p(D_1, D_2; \mathbf{r}) D_1 D_2. 
\label{eq:integral}
\end{equation}
Now, we make use of the invariance of the probability measures,
\[
dD_1 dD_2 p_D(D_1, D_2, \mathbf{r}) = d \widehat{G}_1 d\widehat{G}_2 p_G(\widehat{G}_1, \widehat{G}_2; \mathbf{r}),
\]
with
\[
p_G(\widehat{G}_1, \widehat{G}_2 ; \mathbf{r}) = \frac{1}{2 \pi \sqrt{1 - \rho(\mathbf{r})^2}}  \exp \left[ - \frac{ \widehat{G}_1^2 + \widehat{G}_2^2 - 2 \widehat{G}_1 \widehat{G}_ 2 \rho(\mathbf{r})}{2(1 - \rho(\mathbf{r})^2)} \right]
\]
being the bivariate distribution of the correlated Gaussian field used in the first stage of construction of the diffusivity landscape, with $\rho(\mathbf{r})$ being the correlation function of this field given by Eq.~(\ref{eq:gauss_corr}).
We note that the $\mathbf{r }$-dependence in this expression is fully due to the one of the correlation function $\rho(\mathbf{r})$, and concentrate only on this $\rho$-dependence, introducing the function $g(\widehat{G}_1, \widehat{G}_2 ; \rho) =
p(\widehat{G}_1, \widehat{G}_2 ; \mathbf{r})$.
Now, we can  write Eq.~(\ref{eq:integral}) as
\begin{equation}
\langle D_1 D_2 \rangle  = \int_{-\infty}^{\infty} \int_{\infty}^{\infty} d\widehat{G}_1 d\widehat{G}_2 g(\widehat{G}_1, \widehat{G}_2; \rho) f(\widehat{G}_1) f(\widehat{G}_2),
\label{eq:integral_g}
\end{equation}
with $f(\widehat{G})$ being the function that transforms the correlated Gaussian field into the diffusivity landscape, Eq.~(\ref{eq:change_var}). %), namely
%\begin{equation}
%f(\widehat{G}) = F^{-1}_{\beta} \left\{ \frac{1}{2} \left[ 1 - \text{erf} \left( \frac{\widehat{G}}{\sqrt{2}} \right) \right] \right\}.
%\label{eq:Transf}
%\end{equation}
Note that according to Eqs. (\ref{eq:integral_g}) and (\ref{eq:change_var}) the value of the function $\langle D_1 D_2 \rangle$ is a function of $\rho$ only, and therefore passing to the correlation function of diffusivity landscape, 
which differs from $\langle D(\mathbf{0}) D(\mathbf{r}) \rangle$ by shift and rescaling, we see that $\xi(\mathbf{r}) = \xi[\rho(\mathbf{r})]$, and the dependence $\xi(\rho)$ is not influenced by a particular shape of the correlation 
function of the Gaussian landscape. Thus, the transformation from the Gaussian field to a Gamma-distributed landscape corresponds to a pointwise transformation of their correlation functions.  This property will be used several times.

The integration in Eq.~(\ref{eq:integral_g}) can only be performed numerically. However, one can still find an analytical approximation to this integral. We begin by Taylor expanding the function $f(\widehat{G})$ around zero up to the fourth order:
\[
f(\widehat{G}) \approx a_0 + a_1 \widehat{G} + a_2 \widehat{G}^2 + a_3 \widehat{G}^3 + a_4 \widehat{G}^4 + O(\widehat{G}^5),
\]
with $a_0 = 1.4505$, $a_1 = 0.9704$, $a_2 = 0.2194$, $a_3 = 0.0130$ and $a_4 = 0.0011$ for the values of parameters used. The coefficients correspond to the numerical evaluation of the analytical expressions of the corresponding derivatives of $f$ which is easily done with Mathematica. This last expression can now be used to compute the integral in Eq.~(\ref{eq:integral_g}) as a function of $\rho$, since the corresponding integral reduces to the sum of moments of a bivariate Gaussian weighted with different prefactors. 
Keeping contributions up to the fourth order in $\rho$ we find 
\[
\langle D(\mathbf{0}) D(\mathbf{r}) \rangle \approx b_0 + b_1 \rho + b_2 \rho^2 + b_3 \rho^3 + b_4 \rho^4 + O(\rho^5),
\]
with $b_0 = 2.77717$, $b_1 = 1.01867$, $b_2 = 0.09043$, $b_3 = 0.00101$ and $b_4 = 0.00003$. The first coefficient ($b_0$) is equal to $\overline{D}^2$, therefore it vanishes when plugging
back into Eq.~(\ref{eq:DL_corr}). Moreover, since the coefficients $b_3$ and $b_4$ are small compared
to $b_1$ and $b_2$, they can be neglected. Under this approximation we get $C_{DD}(\mathbf{r}) \approx \xi[\rho(\mathbf{r})]$ with the function
\begin{equation}
 \xi(\rho) =\frac{b_1 \rho + b_2 \rho^2}{b_1 + b_2} = c_1 \rho + c_2 \rho^2.
 \label{eq:Dir}
\end{equation}
with $c_1 = 0.918465$ and $c_2 = 0.081535$. 
This simple quadratic approximation has the relative accuracy better than $0.0005$ in the whole domain $0\leq \rho \leq 1$ as compared to the result of high-precision numerical integration. 

The transformation $\xi(\rho)$ is invertible, and gives therefore the possibility to construct a Gaussian filed whose probability transformation would produce a Gamma-field with a desired two-point
correlation function. We will use this possibility in what follows, when considering the correlated CTRW scheme in Sec. \ref{sec:CorrCTRW}. The inverse transformation is given by the solution of 
the quadratic equation, giving the inverse function 
\begin{equation}
 \rho(\xi) = \sqrt{\left(\frac{c_1}{2 c_2}\right)^2 + \frac{\xi}{2c_2}} - \frac{c_1}{2 c_2}.
 \label{eq:Inv}
\end{equation}

Substituting the expression for $\rho(\mathbf{r})$, Eq.~(\ref{eq:gauss_corr}), into Eq.~(\ref{eq:Dir}) we get the approximation for the correlation function of the diffusivity landscape: 
\begin{equation}
C_{DD}(\mathbf{r}) \approx c_1 \exp \left( - \frac{\mathbf{r}^2}{2\lambda^2} \right) + c_2 \exp \left( - \frac{\mathbf{r}^2}{\lambda^2} \right).
\label{eq:DL_corr_approx}
\end{equation}
Now we can use this approximation to find the correlation function of the diffusivity landscape along the trajectories, or in other words, as a function of the number of steps $C_{DD}(n)$. To do so, Eq.~(\ref{eq:DL_corr_approx}) has to be averaged using the PDF $f(\mathbf{r} \arrowvert n)$ of the displacements given the number of steps $n$:
\begin{equation}
C_{DD}(n) = \int d\mathbf{r} \; C_{DD}(\mathbf{r}) \; f(\mathbf{r} \arrowvert n).
\label{eq:DL_corr_traj}
\end{equation}

\begin{figure}[h!]
    \centering
    \def\svgwidth{0.8\columnwidth}
\begingroup%
  \makeatletter%
  \providecommand\color[2][]{%
    \errmessage{(Inkscape) Color is used for the text in Inkscape, but the package 'color.sty' is not loaded}%
    \renewcommand\color[2][]{}%
  }%
  \providecommand\transparent[1]{%
    \errmessage{(Inkscape) Transparency is used (non-zero) for the text in Inkscape, but the package 'transparent.sty' is not loaded}%
    \renewcommand\transparent[1]{}%
  }%
  \providecommand\rotatebox[2]{#2}%
  \newcommand*\fsize{\dimexpr\f@size pt\relax}%
  \newcommand*\lineheight[1]{\fontsize{\fsize}{#1\fsize}\selectfont}%
  \ifx\svgwidth\undefined%
    \setlength{\unitlength}{792bp}%
    \ifx\svgscale\undefined%
      \relax%
    \else%
      \setlength{\unitlength}{\unitlength * \real{\svgscale}}%
    \fi%
  \else%
    \setlength{\unitlength}{\svgwidth}%
  \fi%
  \global\let\svgwidth\undefined%
  \global\let\svgscale\undefined%
  \makeatother%
  \begin{picture}(1,0.77272727)%
    \lineheight{1}%
    \setlength\tabcolsep{0pt}%
    \put(0,0){\includegraphics[width=\unitlength,page=1]{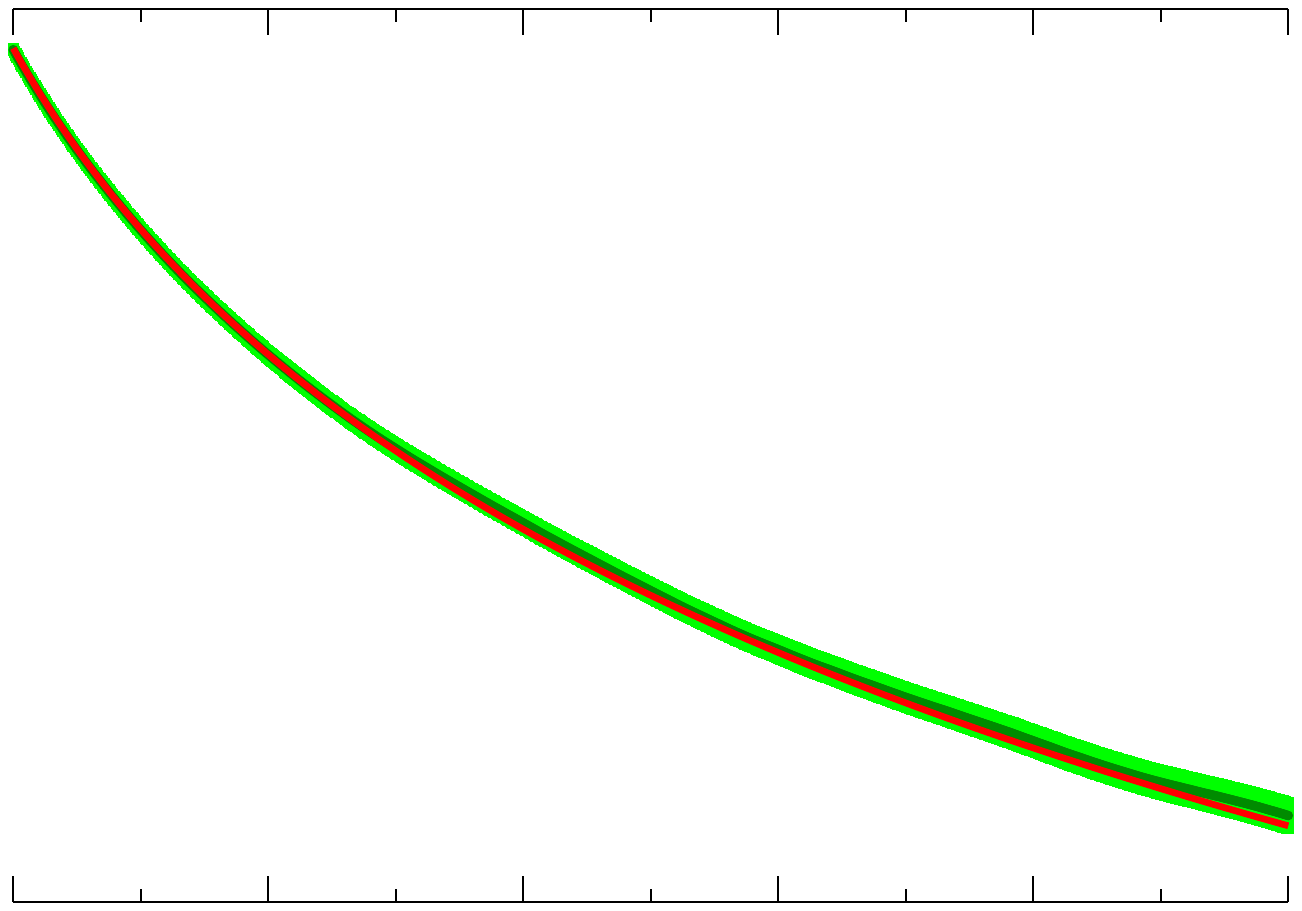}}%
    \put(0.11085859,0.08429293){\makebox(0,0)[lt]{\lineheight{1.25}\smash{\begin{tabular}[t]{l}$0$\end{tabular}}}}%
    \put(0.26467172,0.08429293){\makebox(0,0)[lt]{\lineheight{1.25}\smash{\begin{tabular}[t]{l}$2$\end{tabular}}}}%
    \put(0.41921717,0.08429293){\makebox(0,0)[lt]{\lineheight{1.25}\smash{\begin{tabular}[t]{l}$4$\end{tabular}}}}%
    \put(0.57376263,0.0830303){\makebox(0,0)[lt]{\lineheight{1.25}\smash{\begin{tabular}[t]{l}$6$\end{tabular}}}}%
    \put(0.72830808,0.08429293){\makebox(0,0)[lt]{\lineheight{1.25}\smash{\begin{tabular}[t]{l}$8$\end{tabular}}}}%
    \put(0.87717172,0.0830303){\makebox(0,0)[lt]{\lineheight{1.25}\smash{\begin{tabular}[t]{l}$10$\end{tabular}}}}%
    \put(0.49722222,0.07520202){\makebox(0,0)[lt]{\lineheight{1.25}\smash{\begin{tabular}[t]{l}$n$\end{tabular}}}}%
    \put(0.90717172,0.1230303){\makebox(0,0)[lt]{\lineheight{1.25}\smash{\begin{tabular}[t]{l}$\times 10^{2}$\end{tabular}}}}%
    \put(0,0){\includegraphics[width=\unitlength,page=2]{DD_corr.pdf}}%
    \put(0.03409091,0.10833333){\makebox(0,0)[lt]{\lineheight{1.25}\smash{\begin{tabular}[t]{l}$-2.0$\end{tabular}}}}%
    \put(0.03409091,0.21136364){\makebox(0,0)[lt]{\lineheight{1.25}\smash{\begin{tabular}[t]{l}$-1.6$\end{tabular}}}}%
    \put(0.03409091,0.31439394){\makebox(0,0)[lt]{\lineheight{1.25}\smash{\begin{tabular}[t]{l}$-1.2$\end{tabular}}}}%
    \put(0.03409091,0.41742424){\makebox(0,0)[lt]{\lineheight{1.25}\smash{\begin{tabular}[t]{l}$-0.8$\end{tabular}}}}%
    \put(0.03409091,0.52045455){\makebox(0,0)[lt]{\lineheight{1.25}\smash{\begin{tabular}[t]{l}$-0.4$\end{tabular}}}}%
    \put(0.06166667,0.62348485){\makebox(0,0)[lt]{\lineheight{1.25}\smash{\begin{tabular}[t]{l}$0.0$\end{tabular}}}}%
    \put(0.01636364,0.32121212){\rotatebox{90}{\makebox(0,0)[lt]{\lineheight{1.25}\smash{\begin{tabular}[t]{l}$\ln C_{DD}(n)$\end{tabular}}}}}%
    \put(0,0){\includegraphics[width=\unitlength,page=3]{DD_corr.pdf}}%
    \put(0.70235795,0.59530303){\makebox(0,0)[lt]{\lineheight{1.25}\smash{\begin{tabular}[t]{l}\tiny Sim\end{tabular}}}}%
    \put(0,0){\includegraphics[width=\unitlength,page=4]{DD_corr.pdf}}%
    \put(0.70235795,0.57242424){\makebox(0,0)[lt]{\lineheight{1.25}\smash{\begin{tabular}[t]{l}\tiny Eq.(\ref{eq:DD_corr_n})\end{tabular}}}}%
    \put(0,0){\includegraphics[width=\unitlength,page=5]{DD_corr.pdf}}%
  \end{picture}%
\endgroup%
    \caption{Correlation function $C_{DD}(n)$ as a function of the number of steps for the two-dimensional case with $\lambda = 10$ and $D_0 = 1$. We compare the numerical results obtained in simulations of particle diffusion in the diffusivity landscapes (green line), with the approximation Eq.~(\ref{eq:DD_corr_n}) (red line). The standard errors of the mean (SEM) are represented by the light green area. Excellent agreement is observed in the whole range of the steps' numbers.}
    \label{fig:DD_corr}
\end{figure}

Given that the spatial part of the motion is a two-dimensional simple random walk, the PDF $f(\mathbf{r} \arrowvert n)$ can be safely approximated by a two-dimensional Gaussian distribution %of the form \cite{Klafter2011}
\begin{equation}
f(\mathbf{r} \arrowvert n) = \left(\frac{d}{2 \pi a^2 n} \right)^{\frac{d}{2}} \exp \left( - \frac{d \, \mathbf{r}^2}{2 a^2 n} \right) = \frac{1}{2 \pi \sigma^2 n} \exp \left( - \frac{\mathbf{r}^2}{2 \sigma^2 n} \right),
\label{eq:PDF_dis_n}
\end{equation}
with $d=2$, $a=1$ and, respectively, $\sigma^2 = 1/2$. Within this approximation, Eq.~(\ref{eq:DL_corr_traj}) takes the form
\begin{equation}
C_{DD}(n) \approx  c_1 \left( 1 + \frac{n}{2 \lambda^2} \right)^{-1} + c_2 \left( 1 + \frac{n}{\lambda^2} \right)^{-1}.
\label{eq:DD_corr_n}
\end{equation}
Figure~\ref{fig:DD_corr} shows a comparison between this approximate expression and the result from simulations of particle diffusion on the diffusivity landscape. Excellent agreement is observed in the whole range of steps' numbers. Note that the correlation function of diffusion coefficients is extremely long-ranged.

%---------------------------------------------------------------
%Subection: Correlated CTRW
%---------------------------------------------------------------

\subsection{Correlated CTRW \label{sec:CorrCTRW}}

%\color{red}
Let us now use the correlation function of the diffusivity values along the trajectories, Eq.~(\ref{eq:DD_corr_n}), to construct a time-correlated CTRW scheme. The process of generating correlated waiting times is similar to the one used for generating the landscape. Starting from values of $\xi(n) = C_{DD}(n)$ given by Eq.~(\ref{eq:DD_corr_n}) we use Eq.~(\ref{eq:Inv}) to obtain the correlation function $\rho(n)$ of a Gaussian vector which then will be transformed to the one of diffusivity values 
and finally into waiting times along the trajectory.

We proceed by generating a one-dimensional uncorrelated Gaussian vector $g_i$ by assigning to each entry of the vector a random number drawn from a Gaussian distribution with zero mean and unit variance. Then, using the Fourier filtering method \cite{Toral2014}, we generate a correlated Gaussian vector $\widehat{g}_i$ with correlation function $\rho(n)$ with $n = \arrowvert i - j \arrowvert$.  Using the probability transformation, Eq.~(\ref{eq:change_var}), we transform this correlated Gaussian vector into a one-dimensional array of diffusion coefficients $\mathcal{D}_i$ with the desired correlation function $C_{DD}(n) = \xi(n)$. The array of correlated diffusion coefficients $\mathcal{D}_i$ is then used to generate waiting times of our CTRW scheme by drawing random numbers $t_i$ from an exponential waiting time distribution $\psi(t \arrowvert \mathcal{D}_i) = 4 \mathcal{D}_i \exp(- 4 \mathcal{D}_i t)$. In each realization of the process, one repeats the procedure until obtaining such a number $n'$ of drawn elements that the sum of the first $n^{\prime}$ elements does not exceed $t_{max}$ but the sum of the first $n^{\prime} + 1$  does. The number of elements $n^{\prime}$ is then the number of steps performed by a walker until $t_{max}$. The PDF of displacements for this correlated CTRW can be estimated by the average
\[
p(\mathbf{r},t_{max}) = \langle f(\mathbf{r}\arrowvert n^{\prime}) \rangle_{n^{\prime}},
\]
with $f(\mathbf{r}\arrowvert n)$ being the PDF of displacements for a given number of steps, Eq.~(\ref{eq:PDF_dis_n}), weighted with the waiting time of the first step. 

\begin{figure}[h!]
    \def\svgwidth{0.5\columnwidth}
\begingroup%
  \makeatletter%
  \providecommand\color[2][]{%
    \errmessage{(Inkscape) Color is used for the text in Inkscape, but the package 'color.sty' is not loaded}%
    \renewcommand\color[2][]{}%
  }%
  \providecommand\transparent[1]{%
    \errmessage{(Inkscape) Transparency is used (non-zero) for the text in Inkscape, but the package 'transparent.sty' is not loaded}%
    \renewcommand\transparent[1]{}%
  }%
  \providecommand\rotatebox[2]{#2}%
  \newcommand*\fsize{\dimexpr\f@size pt\relax}%
  \newcommand*\lineheight[1]{\fontsize{\fsize}{#1\fsize}\selectfont}%
  \ifx\svgwidth\undefined%
    \setlength{\unitlength}{424.00089407bp}%
    \ifx\svgscale\undefined%
      \relax%
    \else%
      \setlength{\unitlength}{\unitlength * \real{\svgscale}}%
    \fi%
  \else%
    \setlength{\unitlength}{\svgwidth}%
  \fi%
  \global\let\svgwidth\undefined%
  \global\let\svgscale\undefined%
  \makeatother%
  \begin{picture}(1,0.76650943)%
    \lineheight{1}%
    \setlength\tabcolsep{0pt}%
    \put(0,0){\includegraphics[width=\unitlength,page=1]{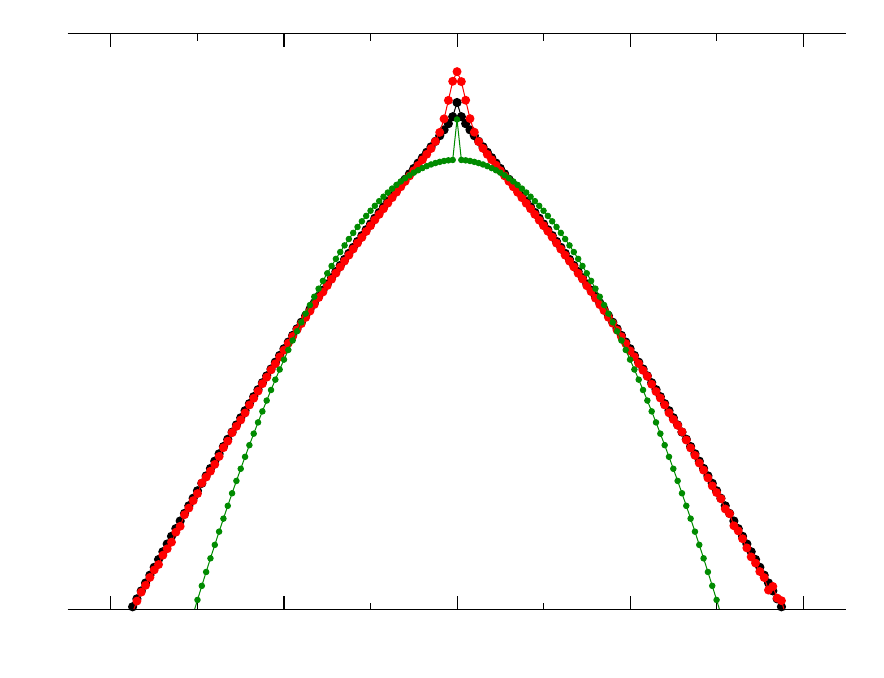}}%
    \put(0.09736769,0.03983491){\makebox(0,0)[lt]{\lineheight{1.25}\smash{\begin{tabular}[t]{l}\tiny $-8$\end{tabular}}}}%
    \put(0.2932533,0.03983491){\makebox(0,0)[lt]{\lineheight{1.25}\smash{\begin{tabular}[t]{l}\tiny $-4$\end{tabular}}}}%
    \put(0.51267689,0.03983491){\makebox(0,0)[lt]{\lineheight{1.25}\smash{\begin{tabular}[t]{l}\tiny $0$\end{tabular}}}}%
    \put(0.70856274,0.03983491){\makebox(0,0)[lt]{\lineheight{1.25}\smash{\begin{tabular}[t]{l}\tiny $4$\end{tabular}}}}%
    \put(0.90444835,0.03983491){\makebox(0,0)[lt]{\lineheight{1.25}\smash{\begin{tabular}[t]{l}\tiny $8$\end{tabular}}}}%
    \put(0.60913915,0.03301887){\makebox(0,0)[lt]{\lineheight{1.25}\smash{\begin{tabular}[t]{l}\footnotesize $\xi$\end{tabular}}}}%
    \put(0,0){\includegraphics[width=\unitlength,page=2]{ctrw_100.pdf}}%
    \put(-0.02596698,0.19580189){\makebox(0,0)[lt]{\lineheight{1.25}\smash{\begin{tabular}[t]{l}\tiny $10^{-4}$\end{tabular}}}}%
    \put(-0.02596698,0.45641509){\makebox(0,0)[lt]{\lineheight{1.25}\smash{\begin{tabular}[t]{l}\tiny $10^{-2}$\end{tabular}}}}%
    \put(0.01068396,0.7170283){\makebox(0,0)[lt]{\lineheight{1.25}\smash{\begin{tabular}[t]{l}\tiny $10^{0}$\end{tabular}}}}%
    \put(-0.01358491,0.38119104){\rotatebox{0}{\makebox(0,0)[lt]{\lineheight{1.25}\smash{\begin{tabular}[t]{l} \footnotesize $q(\xi)$\end{tabular}}}}}%
    \put(0,0){\includegraphics[width=\unitlength,page=3]{ctrw_100.pdf}}%
    \put(0.39772406,0.11015566){\makebox(0,0)[lt]{\lineheight{1.25}\smash{\begin{tabular}[t]{l}\tiny $-1$\end{tabular}}}}%
    \put(0.52034198,0.11015566){\makebox(0,0)[lt]{\lineheight{1.25}\smash{\begin{tabular}[t]{l}\tiny $0$\end{tabular}}}}%
    \put(0.62942217,0.11015566){\makebox(0,0)[lt]{\lineheight{1.25}\smash{\begin{tabular}[t]{l}\tiny $1$\end{tabular}}}}%
    \put(0,0){\includegraphics[width=\unitlength,page=4]{ctrw_100.pdf}}%
    \put(0.28757075,0.16753703){\makebox(0,0)[lt]{\lineheight{1.25}\smash{\begin{tabular}[t]{l}\tiny $10^{-1}$\end{tabular}}}}%
    \put(0,0){\includegraphics[width=\unitlength,page=5]{ctrw_100.pdf}}%
    \put(0.69467007,0.63934847){\makebox(0,0)[lt]{\lineheight{1.25}\smash{\begin{tabular}[t]{l}\footnotesize $(a) \; t = 10^2$\end{tabular}}}}%
  \end{picture}%
\endgroup%
    \def\svgwidth{0.5\columnwidth}
\begingroup%
  \makeatletter%
  \providecommand\color[2][]{%
    \errmessage{(Inkscape) Color is used for the text in Inkscape, but the package 'color.sty' is not loaded}%
    \renewcommand\color[2][]{}%
  }%
  \providecommand\transparent[1]{%
    \errmessage{(Inkscape) Transparency is used (non-zero) for the text in Inkscape, but the package 'transparent.sty' is not loaded}%
    \renewcommand\transparent[1]{}%
  }%
  \providecommand\rotatebox[2]{#2}%
  \newcommand*\fsize{\dimexpr\f@size pt\relax}%
  \newcommand*\lineheight[1]{\fontsize{\fsize}{#1\fsize}\selectfont}%
  \ifx\svgwidth\undefined%
    \setlength{\unitlength}{424.00089407bp}%
    \ifx\svgscale\undefined%
      \relax%
    \else%
      \setlength{\unitlength}{\unitlength * \real{\svgscale}}%
    \fi%
  \else%
    \setlength{\unitlength}{\svgwidth}%
  \fi%
  \global\let\svgwidth\undefined%
  \global\let\svgscale\undefined%
  \makeatother%
  \begin{picture}(1,0.76650943)%
    \lineheight{1}%
    \setlength\tabcolsep{0pt}%
    \put(0,0){\includegraphics[width=\unitlength,page=1]{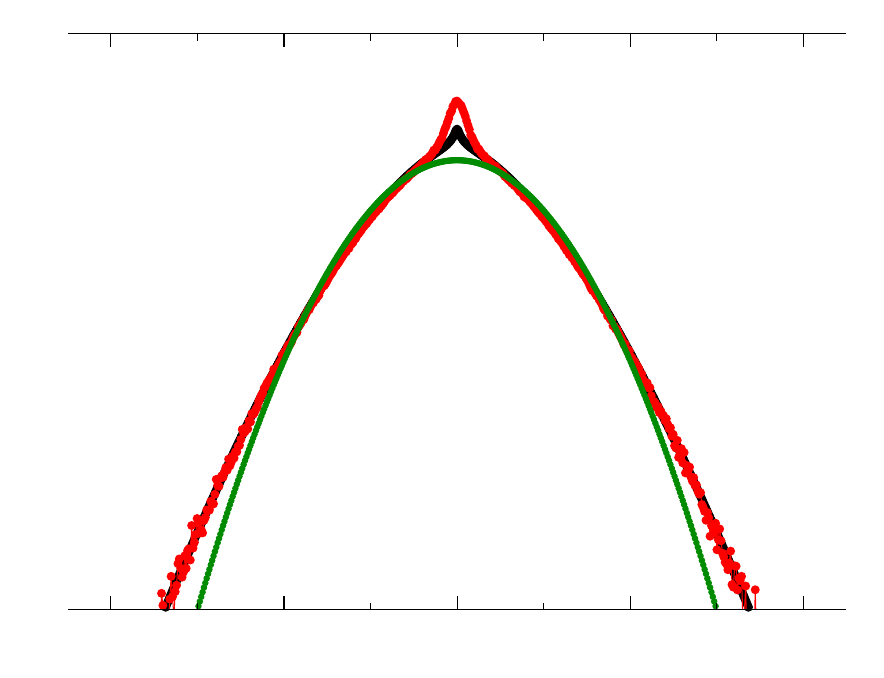}}%
    \put(0.09736769,0.03983491){\makebox(0,0)[lt]{\lineheight{1.25}\smash{\begin{tabular}[t]{l}\tiny $-8$\end{tabular}}}}%
    \put(0.2932533,0.03983491){\makebox(0,0)[lt]{\lineheight{1.25}\smash{\begin{tabular}[t]{l}\tiny $-4$\end{tabular}}}}%
    \put(0.51267689,0.03983491){\makebox(0,0)[lt]{\lineheight{1.25}\smash{\begin{tabular}[t]{l}\tiny $0$\end{tabular}}}}%
    \put(0.70856274,0.03983491){\makebox(0,0)[lt]{\lineheight{1.25}\smash{\begin{tabular}[t]{l}\tiny $4$\end{tabular}}}}%
    \put(0.90444835,0.03983491){\makebox(0,0)[lt]{\lineheight{1.25}\smash{\begin{tabular}[t]{l}\tiny $8$\end{tabular}}}}%
    \put(0.60913915,0.03301887){\makebox(0,0)[lt]{\lineheight{1.25}\smash{\begin{tabular}[t]{l}\footnotesize  $\xi$\end{tabular}}}}%
    \put(0,0){\includegraphics[width=\unitlength,page=2]{ctrw_1000.pdf}}%
    \put(-0.02596698,0.19580189){\makebox(0,0)[lt]{\lineheight{1.25}\smash{\begin{tabular}[t]{l}\tiny $10^{-4}$\end{tabular}}}}%
    \put(-0.02596698,0.45641509){\makebox(0,0)[lt]{\lineheight{1.25}\smash{\begin{tabular}[t]{l}\tiny $10^{-2}$\end{tabular}}}}%
    \put(0.01068396,0.7170283){\makebox(0,0)[lt]{\lineheight{1.25}\smash{\begin{tabular}[t]{l}\tiny $10^{0}$\end{tabular}}}}%
    \put(-0.01358491,0.38119104){\rotatebox{0}{\makebox(0,0)[lt]{\lineheight{1.25}\smash{\begin{tabular}[t]{l}\footnotesize  $q(\xi)$\end{tabular}}}}}%
    \put(0,0){\includegraphics[width=\unitlength,page=3]{ctrw_1000.pdf}}%
    \put(0.39772406,0.11015566){\makebox(0,0)[lt]{\lineheight{1.25}\smash{\begin{tabular}[t]{l}\tiny $-1$\end{tabular}}}}%
    \put(0.52034198,0.11015566){\makebox(0,0)[lt]{\lineheight{1.25}\smash{\begin{tabular}[t]{l}\tiny $0$\end{tabular}}}}%
    \put(0.62942217,0.11015566){\makebox(0,0)[lt]{\lineheight{1.25}\smash{\begin{tabular}[t]{l}\tiny $1$\end{tabular}}}}%
    \put(0,0){\includegraphics[width=\unitlength,page=4]{ctrw_1000.pdf}}%
    \put(0.28757075,0.20857712){\makebox(0,0)[lt]{\lineheight{1.25}\smash{\begin{tabular}[t]{l}\tiny $10^{-1}$\end{tabular}}}}%
    \put(0,0){\includegraphics[width=\unitlength,page=5]{ctrw_1000.pdf}}%
    \put(0.69467007,0.63934847){\makebox(0,0)[lt]{\lineheight{1.25}\smash{\begin{tabular}[t]{l}\footnotesize $(b) \; t = 10^3$\end{tabular}}}}%
  \end{picture}%
\endgroup%
    \caption{A comparison of the one-dimensional cut of the PDF $q(\xi) = p(x,0) t$ of rescaled displacements $\xi = x / \sqrt{t}$ for the decoupled particle (red) and the correlated CTRW (black) for $t=10^2$ and
    $10^3$ when the differences between the behaviors are considerable. The data for the decoupled particles are the same as in Figure~\ref{fig:fake_real}, the result for coupled CTRW correspond to $8 \cdot 10 ^6$ independent
    realizations, see text for details. The green dots show the results for an uncorrelated CTRW model as given by Eqs.~(\ref{eq:CTRW_wtd}) and (\ref{eq:CTRW_wtd_1}). }
    \label{fig:fake_ctrw}
\end{figure}

Figure~\ref{fig:fake_ctrw} shows the resulting PDF for two different times $t_{max}=10^2$, and $10^3$, one time per panel. Each panel presents a comparison between the PDF of the decoupled particle and that in the correlated CTRW. The PDFs for decoupled particles are the same as the ones in Figure~\ref{fig:fake_real} for the corresponding times. For the correlated CTRW, each PDF corresponds to the average over $8 \times 10^6$ different realizations of the correlated array of diffusion coefficients $\mathcal{D}_i$, constructed with $\lambda = 10$. As one can see, both PDFs are indistinguishable in their wings, and both present a central peak which, instead of narrowing, flattens out. However, the shapes of the peaks in both cases are significantly different. We note that the uncorrelated CTRW model shows a very different behavior in the wing (its convergence to a Gaussian is much faster) and does not show any peak except for some remains of initial condition at a shorter time. 

It is worth mentioning that to generate the PDFs of the correlated CTRW we have used an extremely high number of realizations, namely $8 \times 10^6$; in our simulations this number was subdivided into five independent runs, and the results were 
both considered separately, and pooled for the plot in Figure~\ref{fig:fake_ctrw}. The analysis of the subsets shows that the height of the peak in different sets of $1.6 \times 10^6$ realizations still fluctuates considerably, so that 
this height is dominated by rare events, while both in the initial model and in the decoupled variant thereof the behavior in the peak may be considered as much more typical.

Since the approximations used to construct the correlated CTRW are quite accurate and the number of realizations is high enough to guarantee sufficiently good statistics, the differences suggest that our correlated model 
fails to capture important details of temporal correlations. Since serial correlations along the trajectories are reproduced correctly, one can conclude that these are the higher-oder correlations that play a key role 
in the development of the central peak but are of minor importance in the wings.

%---------------------------------------------------------------
%Section: The checkerboard model
%---------------------------------------------------------------

\section{The checkerboard model \label{sec:checkerboard}}

\begin{figure}[h!]
    \centering
    \def\svgwidth{\columnwidth}
\begingroup%
  \makeatletter%
  \providecommand\color[2][]{%
    \errmessage{(Inkscape) Color is used for the text in Inkscape, but the package 'color.sty' is not loaded}%
    \renewcommand\color[2][]{}%
  }%
  \providecommand\transparent[1]{%
    \errmessage{(Inkscape) Transparency is used (non-zero) for the text in Inkscape, but the package 'transparent.sty' is not loaded}%
    \renewcommand\transparent[1]{}%
  }%
  \providecommand\rotatebox[2]{#2}%
  \newcommand*\fsize{\dimexpr\f@size pt\relax}%
  \newcommand*\lineheight[1]{\fontsize{\fsize}{#1\fsize}\selectfont}%
  \ifx\svgwidth\undefined%
    \setlength{\unitlength}{460.79998848bp}%
    \ifx\svgscale\undefined%
      \relax%
    \else%
      \setlength{\unitlength}{\unitlength * \real{\svgscale}}%
    \fi%
  \else%
    \setlength{\unitlength}{\svgwidth}%
  \fi%
  \global\let\svgwidth\undefined%
  \global\let\svgscale\undefined%
  \makeatother%
  \begin{picture}(1,0.75)%
    \lineheight{1}%
    \setlength\tabcolsep{0pt}%
    \put(0,0){\includegraphics[width=\unitlength,page=1]{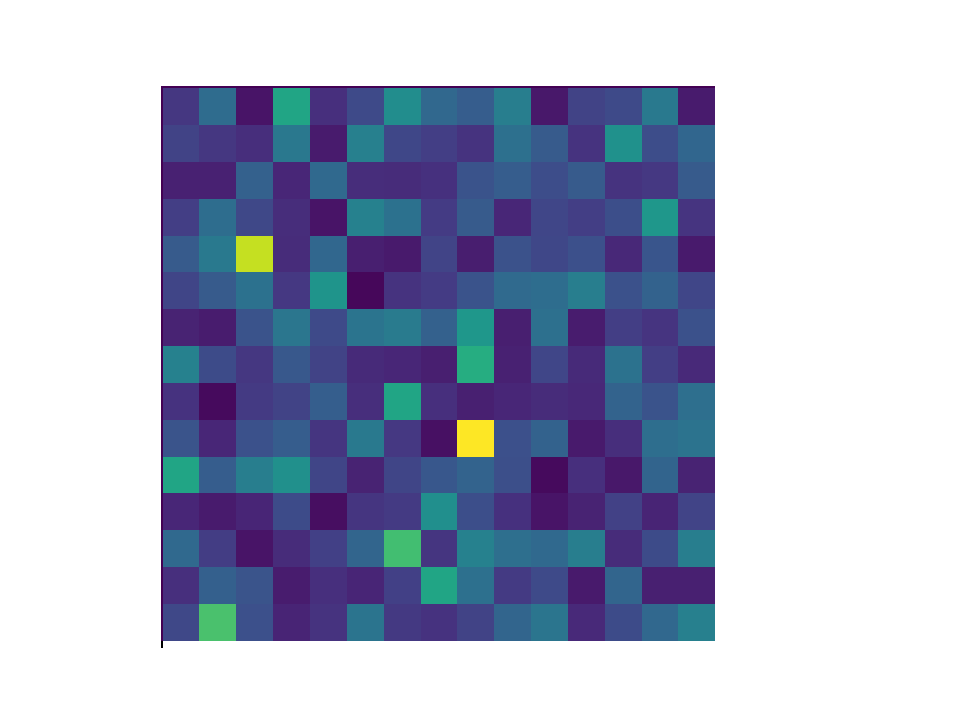}}%
    \put(0.16155924,0.05082964){\color[rgb]{0,0,0}\makebox(0,0)[lt]{\lineheight{1.25}\smash{\begin{tabular}[t]{l}\small $0$\end{tabular}}}}%
    \put(0,0){\includegraphics[width=\unitlength,page=2]{checkboard.pdf}}%
    \put(0.33961806,0.05082964){\color[rgb]{0,0,0}\makebox(0,0)[lt]{\lineheight{1.25}\smash{\begin{tabular}[t]{l}\small $100$\end{tabular}}}}%
    \put(0.43992824,0.04955162){\color[rgb]{0,0,0}\makebox(0,0)[lt]{\lineheight{1.25}\smash{\begin{tabular}[t]{l}$x$\end{tabular}}}}%
    \put(0,0){\includegraphics[width=\unitlength,page=3]{checkboard.pdf}}%
    \put(0.53148003,0.05082964){\color[rgb]{0,0,0}\makebox(0,0)[lt]{\lineheight{1.25}\smash{\begin{tabular}[t]{l}\small $200$\end{tabular}}}}%
    \put(0,0){\includegraphics[width=\unitlength,page=4]{checkboard.pdf}}%
    \put(0.72333984,0.05082964){\color[rgb]{0,0,0}\makebox(0,0)[lt]{\lineheight{1.25}\smash{\begin{tabular}[t]{l}\small $300$\end{tabular}}}}%
    \put(0,0){\includegraphics[width=\unitlength,page=5]{checkboard.pdf}}%
    \put(0.13850825,0.65305556){\color[rgb]{0,0,0}\makebox(0,0)[lt]{\lineheight{1.25}\smash{\begin{tabular}[t]{l}\small $0$\end{tabular}}}}%
    \put(0,0){\includegraphics[width=\unitlength,page=6]{checkboard.pdf}}%
    \put(0.1109069,0.46119575){\color[rgb]{0,0,0}\makebox(0,0)[lt]{\lineheight{1.25}\smash{\begin{tabular}[t]{l}\small $100$\end{tabular}}}}%
    \put(0,0){\includegraphics[width=\unitlength,page=7]{checkboard.pdf}}%
    \put(0.1109069,0.26933594){\color[rgb]{0,0,0}\makebox(0,0)[lt]{\lineheight{1.25}\smash{\begin{tabular}[t]{l}\small $200$\end{tabular}}}}%
    \put(0,0){\includegraphics[width=\unitlength,page=8]{checkboard.pdf}}%
    \put(0.1109069,0.07747483){\color[rgb]{0,0,0}\makebox(0,0)[lt]{\lineheight{1.25}\smash{\begin{tabular}[t]{l}\small $300$\end{tabular}}}}%
    \put(0.09771636,0.36482422){\color[rgb]{0,0,0}\rotatebox{0}{\makebox(0,0)[lt]{\lineheight{1.25}\smash{\begin{tabular}[t]{l}$y$\end{tabular}}}}}%
    \put(0,0){\includegraphics[width=\unitlength,page=9]{checkboard.pdf}}%
    \put(0.82781684,0.0742602){\color[rgb]{0,0,0}\makebox(0,0)[lt]{\lineheight{1.25}\smash{\begin{tabular}[t]{l}$0$\end{tabular}}}}%
    \put(0,0){\includegraphics[width=\unitlength,page=10]{checkboard.pdf}}%
    \put(0.82781684,0.15761306){\color[rgb]{0,0,0}\makebox(0,0)[lt]{\lineheight{1.25}\smash{\begin{tabular}[t]{l}$1$\end{tabular}}}}%
    \put(0,0){\includegraphics[width=\unitlength,page=11]{checkboard.pdf}}%
    \put(0.82781684,0.24096571){\color[rgb]{0,0,0}\makebox(0,0)[lt]{\lineheight{1.25}\smash{\begin{tabular}[t]{l}$2$\end{tabular}}}}%
    \put(0,0){\includegraphics[width=\unitlength,page=12]{checkboard.pdf}}%
    \put(0.82781684,0.32431858){\color[rgb]{0,0,0}\makebox(0,0)[lt]{\lineheight{1.25}\smash{\begin{tabular}[t]{l}$3$\end{tabular}}}}%
    \put(0,0){\includegraphics[width=\unitlength,page=13]{checkboard.pdf}}%
    \put(0.82781684,0.40767144){\color[rgb]{0,0,0}\makebox(0,0)[lt]{\lineheight{1.25}\smash{\begin{tabular}[t]{l}$4$\end{tabular}}}}%
    \put(0,0){\includegraphics[width=\unitlength,page=14]{checkboard.pdf}}%
    \put(0.82781684,0.49102648){\color[rgb]{0,0,0}\makebox(0,0)[lt]{\lineheight{1.25}\smash{\begin{tabular}[t]{l}$5$\end{tabular}}}}%
    \put(0,0){\includegraphics[width=\unitlength,page=15]{checkboard.pdf}}%
    \put(0.82781684,0.57437934){\color[rgb]{0,0,0}\makebox(0,0)[lt]{\lineheight{1.25}\smash{\begin{tabular}[t]{l}6\end{tabular}}}}%
    \put(0.83667824,0.65564352){\color[rgb]{0,0,0}\rotatebox{0}{\makebox(0,0)[lt]{\lineheight{1.25}\smash{\begin{tabular}[t]{l}\large $D(x,y)$\end{tabular}}}}}%
  \end{picture}%
\endgroup%
    \caption{A two dimensional realization of the diffusivity landscape $D(\mathbf{r})$ for the checkerboard model. It corresponds to a $300 \times 300$ lattice where each cell has a size of $2 \zeta$ with $\zeta = 10$, and sampled diffusion coefficient is $D_0 = 1$.
    }
    \label{fig:checkerboard_land}
\end{figure}

Let us now go  a few steps back and consider how critical our assumption about the shape of the correlation function $\rho(\mathbf{r})$, Eq.~(\ref{eq:gauss_corr}), is, i.e., what happens if this function is chosen differently. To do so, we consider a DLM with a checkerboard-like diffusivity landscape. On a lattice, a checkerboard-like diffusivity landscape consists of an array of $N \times N$ squares containing $2 \zeta \times 2 \zeta$ lattice points. A constant diffusion coefficient $D^{(i)}$ is assigned to each square. These diffusion coefficients are drawn from the distribution given by Eq.~(\ref{eq:dlm_pD}), the condition needed for the diffusion to be BnG. This choice of diffusivity landscape strongly changes the shape of the correlation function. Moreover, the changes in diffusion coefficients on the borders of the squares are now discontinuous, while the previous diffusivity landscape was assumed to model a smooth situation. In this model, $\zeta$ defines the correlation length of the landscape; to compare to the results of the above DLM, we set $\zeta = \lambda$. Figure~\ref{fig:checkerboard_land} shows one realization of the checkerboard-like diffusivity landscape on a lattice of $300 \times 300$ with $\zeta = 10$ and $D_0 = 1$. 

\begin{figure}[h!]
    \centering
    \def\svgwidth{0.8\columnwidth}
\begingroup%
  \makeatletter%
  \providecommand\color[2][]{%
    \errmessage{(Inkscape) Color is used for the text in Inkscape, but the package 'color.sty' is not loaded}%
    \renewcommand\color[2][]{}%
  }%
  \providecommand\transparent[1]{%
    \errmessage{(Inkscape) Transparency is used (non-zero) for the text in Inkscape, but the package 'transparent.sty' is not loaded}%
    \renewcommand\transparent[1]{}%
  }%
  \providecommand\rotatebox[2]{#2}%
  \newcommand*\fsize{\dimexpr\f@size pt\relax}%
  \newcommand*\lineheight[1]{\fontsize{\fsize}{#1\fsize}\selectfont}%
  \ifx\svgwidth\undefined%
    \setlength{\unitlength}{424.00089407bp}%
    \ifx\svgscale\undefined%
      \relax%
    \else%
      \setlength{\unitlength}{\unitlength * \real{\svgscale}}%
    \fi%
  \else%
    \setlength{\unitlength}{\svgwidth}%
  \fi%
  \global\let\svgwidth\undefined%
  \global\let\svgscale\undefined%
  \makeatother%
  \begin{picture}(1,0.76650943)%
    \lineheight{1}%
    \setlength\tabcolsep{0pt}%
    \put(0,0){\includegraphics[width=\unitlength,page=1]{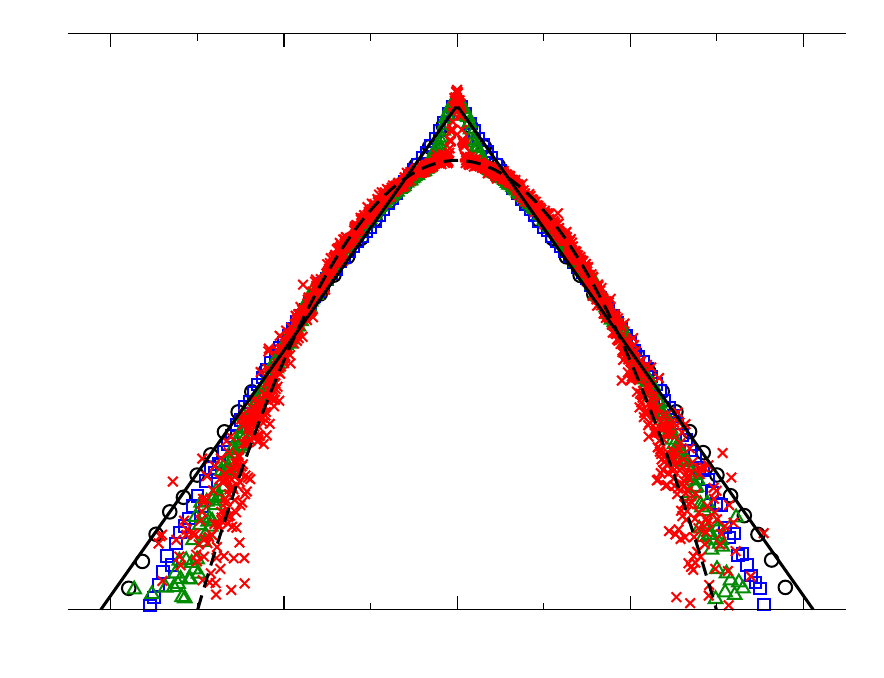}}%
    \put(0.09736769,0.04483491){\makebox(0,0)[lt]{\lineheight{1.25}\smash{\begin{tabular}[t]{l}$-8$\end{tabular}}}}%
    \put(0.2932533,0.04483491){\makebox(0,0)[lt]{\lineheight{1.25}\smash{\begin{tabular}[t]{l}$-4$\end{tabular}}}}%
    \put(0.51267689,0.04483491){\makebox(0,0)[lt]{\lineheight{1.25}\smash{\begin{tabular}[t]{l}$0$\end{tabular}}}}%
    \put(0.70856274,0.04483491){\makebox(0,0)[lt]{\lineheight{1.25}\smash{\begin{tabular}[t]{l}$4$\end{tabular}}}}%
    \put(0.90444835,0.04483491){\makebox(0,0)[lt]{\lineheight{1.25}\smash{\begin{tabular}[t]{l}$8$\end{tabular}}}}%
    \put(0.60913915,0.03301887){\makebox(0,0)[lt]{\lineheight{1.25}\smash{\begin{tabular}[t]{l}$\xi$\end{tabular}}}}%
    \put(0,0){\includegraphics[width=\unitlength,page=2]{checkerboard_pdf.pdf}}%
    \put(-0.00596698,0.19580189){\makebox(0,0)[lt]{\lineheight{1.25}\smash{\begin{tabular}[t]{l}$10^{-4}$\end{tabular}}}}%
    \put(-0.00596698,0.45641509){\makebox(0,0)[lt]{\lineheight{1.25}\smash{\begin{tabular}[t]{l}$10^{-2}$\end{tabular}}}}%
    \put(0.02068396,0.7170283){\makebox(0,0)[lt]{\lineheight{1.25}\smash{\begin{tabular}[t]{l}$10^{0}$\end{tabular}}}}%
    \put(0.01358491,0.38119104){\rotatebox{0}{\makebox(0,0)[lt]{\lineheight{1.25}\smash{\begin{tabular}[t]{l}$q(\xi)$\end{tabular}}}}}%
    \put(0,0){\includegraphics[width=\unitlength,page=3]{checkerboard_pdf.pdf}}%
    \put(0.74884434,0.65863797){\makebox(0,0)[lt]{\lineheight{1.25}\smash{\begin{tabular}[t]{l}\footnotesize $t = 10^1$\end{tabular}}}}%
    \put(0,0){\includegraphics[width=\unitlength,page=4]{checkerboard_pdf.pdf}}%
    \put(0.74884434,0.62915684){\makebox(0,0)[lt]{\lineheight{1.25}\smash{\begin{tabular}[t]{l}\footnotesize $t = 10^2$\end{tabular}}}}%
    \put(0,0){\includegraphics[width=\unitlength,page=5]{checkerboard_pdf.pdf}}%
    \put(0.74884434,0.59967571){\makebox(0,0)[lt]{\lineheight{1.25}\smash{\begin{tabular}[t]{l}\footnotesize $t = 10^3$\end{tabular}}}}%
    \put(0,0){\includegraphics[width=\unitlength,page=6]{checkerboard_pdf.pdf}}%
    \put(0.74884434,0.57019458){\makebox(0,0)[lt]{\lineheight{1.25}\smash{\begin{tabular}[t]{l}\footnotesize $t = 10^4$\end{tabular}}}}%
    \put(0,0){\includegraphics[width=\unitlength,page=7]{checkerboard_pdf.pdf}}%
    \put(0.39772406,0.11615566){\makebox(0,0)[lt]{\lineheight{1.25}\smash{\begin{tabular}[t]{l}\footnotesize $-1$\end{tabular}}}}%
    \put(0.52034198,0.11615566){\makebox(0,0)[lt]{\lineheight{1.25}\smash{\begin{tabular}[t]{l}\footnotesize $0$\end{tabular}}}}%
    \put(0.62942217,0.11615566){\makebox(0,0)[lt]{\lineheight{1.25}\smash{\begin{tabular}[t]{l}\footnotesize $1$\end{tabular}}}}%
    \put(0,0){\includegraphics[width=\unitlength,page=8]{checkerboard_pdf.pdf}}%
    \put(0.32757075,0.20857712){\makebox(0,0)[lt]{\lineheight{1.25}\smash{\begin{tabular}[t]{l}\footnotesize $10^{-1}$\end{tabular}}}}%
    \put(0,0){\includegraphics[width=\unitlength,page=9]{checkerboard_pdf.pdf}}%
  \end{picture}%
\endgroup%
    \caption{A one dimensional cut of the PDF $q(\xi) = p(x,0) t$ of rescaled displacements $\xi = x / \sqrt{t} $ for the diffusion of particles in the checkerboard model. The straight line corresponds to the Laplace distribution, whereas the dotted line corresponds to the Gaussian distribution. The inset shows a close-up of the central part of the distribution exposing the peak.}
    \label{fig:PDF_checkerboard}
\end{figure}

As in the case of the DLM, we perform random walk simulations of particles diffusing on an ensemble of checkerboard-like landscapes, from which the PDF of displacements can then be constructed. Figure~\ref{fig:PDF_checkerboard} shows the time evolution of the PDF averaged over $2 \times 10^4$ different realizations of the landscape, each one using $10^4$ particles. The landscape was constructed with $N = 109$ and $\zeta = 10$, i.e., we consider a lattice of size $2180 \times 2180$.  One can see that the central peak is preserved. Moreover, the transition to the Gaussian limits follows the same type of convergence by its narrowing. This suggests that the form of the correlation function of the diffusivity landscape does not change the overall behavior. A closer look, though, reveals the presence of some discontinuities near the center of the distribution, which are expected from the fact that the diffusivity landscape itself is discontinuous.

%---------------------------------------------------------------
%Section: Conclusions
%---------------------------------------------------------------

\section{Conclusions \label{sec:conclu}}

In this work, we study the diffusivity landscape model (DLM) characterized by a diffusion coefficient slowly varying in space. Under specific conditions, this model leads to a Brownian yet non-Gaussian diffusion, that is, the MSD is linear in time, but the shape of the PDF changes from a Laplace distribution at short times to a Gaussian distribution at long ones. The art of convergence to the Gaussian is quite a peculiar one since the PDF at all times displays a central peak that does not decay with time, but narrows under rescaling. We show that the persistence of the peak is due to strong spatiotemporal correlations introduced by correlations of local diffusion coefficients in space. Destroying the spatiotemporal correlations on the level of single trajectories (by considering a different relaization of steps' directions while keeping the same list of waiting times as for the real motion) lets the peak to lower and to disappear at longer times. 
This kind of behavior is qualitatively reproduced by a correlated CTRW model with serial correlations of waiting times along the trajectory
mimicking the ones observed in simulations. The model however fails to quantitatively reproduce the PDF for the decoupled case, showing a considerably lower peak. We attribute this fact to an important role of higher-order correlations which are 
not reproduced by the model. By considering a different variant of correlated disorder (the checkerboard model) we moreover show that the existence of the peak is insensitive to the exact shape of the correlation function of local diffusivities, and its shape is hardly sensitive to it. 

\bmhead{Acknowledgments}

The work of A. P. P. was financially supported by Doctoral Programmes in Germany funded by the Deutscher Akademischer Austauschdienst (DAAD) (Programme ID 57440921).

% \section*{Declarations}
% 
% 
% \begin{appendices}
% 
% 
% %%=============================================%%
% %% For submissions to Nature Portfolio Journals %%
% %% please use the heading ``Extended Data''.   %%
% %%=============================================%%
% 
% %%=============================================================%%
% %% Sample for another appendix section			       %%
% %%=============================================================%%
% 
% %% \section{Example of another appendix section}\label{secA2}%
% %% Appendices may be used for helpful, supporting or essential material that would otherwise 
% %% clutter, break up or be distracting to the text. Appendices can consist of sections, figures, 
% %% tables and equations etc.
% 
% \end{appendices}

\end{document}